\documentclass[a4paper,11pt]{article}
\pdfoutput=1 

\usepackage{jheppub} 

\usepackage[T1]{fontenc} 

\usepackage{Settings}
\usepackage{blkarray}
\usepackage{todonotes}

\usepackage[colorlinks = true,
            linkcolor = blue,
            urlcolor  = blue,
            citecolor = blue,
            anchorcolor = blue]{hyperref}

\preprint{SAGEX-21-32-E}

\DeclareMathOperator{\Tr}{Tr}
\newcommand*\DAlambert{\mathop{}\!\mathbin\Box}
\usepackage{blindtext}%
\usepackage{etoolbox}%
\let\oldbibliography\bibliography
\renewcommand{\bibliography}[1]{{%
  \let\chapter\section
  \oldbibliography{#1}}}
\title{\boldmath Graviton particle statistics and coherent states from classical scattering amplitudes}

\author{Ruth Britto, Riccardo Gonzo, Guy R.~Jehu}

\affiliation{School of Mathematics and Hamilton Mathematics Institute\\
Trinity College Dublin\\
 Dublin 2, Ireland}

\emailAdd{britto@maths.tcd.ie,gonzo@maths.tcd.ie,jehu@maths.tcd.ie}

\abstract{In the two-body scattering problem in general relativity, we study the final graviton particle distribution using a perturbative approach. We compute the mean, the variance and the factorial moments of the distribution from the expectation value of the graviton number operator in the KMOC formalism. For minimally coupled scalar particles, the leading deviation from the Poissonian distribution is given by the unitarity cut involving the six-point tree amplitude with the emission of two gravitons. We compute this amplitude in two independent ways. First, we use an extension of the Cheung-Remmen parametrization that includes minimally coupled scalars. We then repeat the calculation using on-shell BCFW-like techniques, finding complete agreement. In the classical limit, this amplitude gives a purely quantum contribution, proving that we can describe the final semiclassical radiation state as a coherent state at least up to order $\mathcal{O}(G^4)$ for classical radiative observables. Finally, we give general arguments about why we expect this to hold also at higher orders in perturbation theory.}

\begin{document}

\maketitle

\section{Introduction}

The two-body problem in general relativity has been receiving increased attention since the first detection of gravitational waves. Quantum field theory methods have recently proven to be very useful to understand the perturbative long-distance regime of the two-body dynamics, offering a new perspective for understanding the inspiral phase in the post-Minkowskian (PM) expansion in the spirit of effective field theories (EFT) \cite{Goldberger:2004jt,Porto:2016pyg}. 

On-shell scattering amplitudes techniques, powered by locality, unitarity and double copy, have been used to get compact analytic expressions for the state-of-the-art binary dynamics for spinless pointlike bodies at 3 PM order and partially at 4 PM order \cite{Neill:2013wsa,Cheung:2018wkq,Bern:2019nnu,Bern:2019crd,Bern:2021dqo}. A handful of alternative and complementary approaches have also been developed in recent years. The relativistic eikonal expansion \cite{KoemansCollado:2019ggb,Cristofoli:2020uzm,DiVecchia:2020ymx,DiVecchia:2021ndb,DiVecchia:2021bdo,Bjerrum-Bohr:2021vuf,Bjerrum-Bohr:2021din} and semiclassical worldline tools \cite{Porto:2016pyg,Kalin:2020mvi,Mogull:2020sak,Kalin:2020fhe,Dlapa:2021npj} offer many insights on the binary problem, both at the conceptual and at the practical computational level. Moreover, the formalism can be extended to include both spinning bodies \cite{Bern:2020buy,Aoude:2021oqj} and finite size effects  \cite{Haddad:2020que,Bern:2020uwk} in terms of additional higher-dimensional operators. All these approaches share the need for careful analysis of the classical limit, as done in seminal work by Kosower, Maybee and O'Connell (KMOC) \cite{Kosower:2018adc}. In the conservative case, a  dictionary  has been found \cite{Kalin:2019rwq,Kalin:2019inp} which enables the analytic continuation of observables from hyperbolic-like scattering orbits to bound orbits, which ultimately are of direct relevance to LIGO. 

The dynamics of the binary in the presence of radiation is much less understood compared to the conservative case. This is very important, for example to establish a direct connection with the waveforms \cite{Mougiakakos:2021ckm,Cristofoli:2021vyo,Jakobsen:2021smu,Jakobsen:2021lvp,Bautista:2021inx}. Unitarity dictates that, even at the classical level, observables are IR-finite only when we include both real and virtual radiation, as stressed in \cite{Herrmann:2021lqe,Herrmann:2021tct}. This is crucial to obtain a well-behaved scattering angle at high energies \cite{Amati:1987wq}, as was proven by a direct calculation of radiation reaction effects \cite{Bern:2020gjj,Parra-Martinez:2020dzs,DiVecchia:2021ndb,DiVecchia:2021bdo,Bini:2021gat}. A similar principle holds for less inclusive observables like gravitational energy event shapes \cite{Gonzo:2020xza}.

We would like to understand the exact structure of the final semiclassical state, including classical radiation. Many recent insights, coming both from a pure worldline description \cite{Mogull:2020sak,Laenen:2008gt,Luna:2016idw,White:2011yy} and from a different parametrization of the kinematics in the classical limit \cite{Parra-Martinez:2020dzs,Bern:2020gjj,Brandhuber:2021kpo,Brandhuber:2021eyq}, suggest that we should expect an (eikonal) exponentiation at all orders in the impact parameter space.  The situation is less clear when we allow particle production. Since we expect the description of a classical wave for a pure state to be possible in terms of a single coherent state \cite{HILLERY1985409,Cristofoli:2021vyo}, a naive crossing-symmetry argument suggests that it should also be possible to describe the final radiation in terms of coherent states. A lot of attention has been devoted so far to the soft expansion where coherent states arise naturally from classical currents \cite{Laddha:2018rle,Sahoo:2018lxl,Ciafaloni:2018uwe,Addazi:2019mjh,Manu:2020zxl,Monteiro:2020plf}, but the dynamics of how these states are generated by the scattering process is much less clear.\footnote{See however \cite{Ware:2013zja} for a notable exception at the leading order in the soft expansion.}

In this work, we compute the expectation value of the graviton number operator using the KMOC formalism, and we show how this is connected to unitarity cuts involving amplitudes with gravitons. Similar ideas in a purely off-shell Schwinger-Keldysh formalism have been developed in \cite{Gelis:2006yv,Gelis:2006cr}. Since coherent states correspond to Poissonian distributions at the level of the particle emission, deviations from such structure imply that we cannot represent the final graviton state as a coherent state. We will show that the leading contribution is given by a unitary cut involving the 6-point tree amplitude $\mathcal{A}_6^{(0)}(\phi_A \phi_B \to \phi_A \phi_B h_1 h_2)$. A complementary perspective is provided by the study of the factorization of radiative observables in the classical limit, as discussed in \cite{Cristofoli:2021jas}.

There are various approaches which can be taken to compute the amplitude in question.  Applying a traditional Feynman-diagram method to gravitational theories is notoriously difficult, due to the multiplicity of gauge-dependent vertex rules contributing to the amplitudes. Inspired by the simplicity of on-shell amplitudes methods, Cheung and Remmen~\cite{Cheung:2017kzx} rewrote the Einstein-Hilbert action in a simpler form through the introduction of an auxiliary field, in the spirit of the first-order formalism developed by Deser~\cite{Deser:1969wk}. The result is a set of Feynman rules for pure gravity which can be used to compute graviton amplitudes in a very efficient way \cite{Abreu:2020lyk}. Here, we will extend this construction to matter by adding minimally coupled scalar fields.\footnote{See also \cite{Gomez:2021shh} for a recent approach based on the perturbiner method.} An alternative approach is to use  on-shell BCFW recursion~\cite{Britto2005a}, which has a number of benefits over off-shell approaches coming from Lagrangians. The first is that the only objects needed are exceptionally simple seed amplitudes which form the base point of the recursion. This approach can in principle be used to reproduce all tree amplitudes of a variety of massless QFTs, including Einstein gravity~\cite{Benincasa:2007xk,Arkani-Hamed:2017jhn}. While the introduction of massive matter does not necessarily obstruct the BCFW construction of higher-point amplitudes~\cite{Badger2005}, the method generally relies on having massless external particles whose null momenta are used to construct a linear momentum shift, and further generally requires the absence (or good behavior) of boundary terms in a corresponding complex integral. Recent works have explored the prospect of applying shifts to massive legs, as well as combining the shift with a soft recursion relation to construct higher-point amplitudes in general massive theories~\cite{Falkowski:2020aso,Ballav:2020ese,Ballav:2021ahg,Lazopoulos:2021mna}. Here, we introduce a new shift capable of reproducing the hard collinear factorizations in a mixed gravity-scalar theory. Using this ``equal-mass shift'' we will show that it is possible to obtain a compact form of $\mathcal{A}_5^{(0)}(\phi_A \phi_B \to \phi_A \phi_B h_1 )$, whilst maintaining gauge invariance at every stage in the computation. We then use the standard BCFW shift to compute the leading classical scaling of $\mathcal{A}_6^{(0)}(\phi_A \phi_B \to \phi_A \phi_B h_1 h_2)$, which is one of the main results of the paper. 

A summary of the paper is as follows. In section~\ref{sec:gravstats}, we study the graviton number operator expectation value in the KMOC formalism for the two-body problem, which is given in terms of unitarity cuts involving on-shell amplitudes with graviton emissions. The deviation from the Poissonian statistics is equivalent to a deviation from coherence in the final semiclassical state with radiation. In section~\ref{sec:Feynman}, we lay out the Feynman-diagram computation, establishing the Feynman rules in the extended Cheung-Remmen formalism with minimally coupled massive scalar particles. In section~\ref{sec:onsh}, we repeat the computation of the five- and six-point amplitudes using an on-shell equal-mass shift and BCFW recursion. In section~\ref{sec:classical_limit}, we study the scaling of these amplitudes in the limit $\hbar \to 0$, and we prove that the six-point tree amplitude does not contribute to the total energy emitted with classical gravitational waves. Moreover, assuming coherence, we also establish new relations in the classical limit between unitarity cuts of amplitudes involving an emission of gravitons in the final state. Section~\ref{sec:conclusion} contains our concluding comments. Appendix A discuss the derivation of the KMOC formalism based on the Schwinger-Keldysh approach, and appendix B summarizes the connection between Poissonian statistics and coherent states.

\paragraph{Conventions}
We work in the mostly plus signature, for consistency with the gravity action presented in~\cite{Cheung:2017kzx}. 

\section{Graviton particle statistics from on-shell amplitudes}
\label{sec:gravstats}
In this section, we study the particle statistics distribution of the gravitons emitted in the scattering of a pair of massive point particles of mass $m_A$ and $m_B$ in general relativity, using methods of perturbative QFT. In particular, we relate the expectation value of the graviton number operator to a sum of unitarity cuts involving scattering amplitudes with external gravitons.

\subsection{Graviton emission probabilities in the KMOC formalism}

Let $\bar{P}_n$ be the probability of emitting $n$ gravitons in the scattering of a pair of massive particles as described above. Unitarity implies that $\sum_{n=0}^{\infty} \bar{P}_n = 1$.
In quantum field theory, this statement is equivalent to a completeness relation in the Hilbert space, 
\begin{align}
\label{eq:completeness}
\ket{0}\bra{0} +\sum_{n=1}^{\infty} \frac{1}{n!} \sum_{\sigma_1,...,\sigma_n = \pm} \int \prod_{i=1}^n d \Phi(k_i) \ket{k_1^{\sigma_1} ... k_n^{\sigma_n}}\bra{k_1^{\sigma_1} ... k_n^{\sigma_n}}=1,
\end{align}
where $\ket{k_1^{\sigma_1} ... k_n^{\sigma_n}}\bra{k_1^{\sigma_1} ... k_n^{\sigma_n}}$ is the $n$-graviton particle projector on states with definite momenta $k_1,...,k_n$ and helicities $\sigma_1,...,\sigma_n$, whose values are indicated by the single $+$ and $-$ signs.

We denote the scattering matrix operator by $S$, the momenta of the incoming (resp.\ outgoing) massive scalar particles by $p_1,p_2$ (resp.\ $p_3,p_4$), and the outgoing graviton momenta by $\{k_i\}_{i=1,...,n}$. It is clear that the probability $\bar P_n$ is given by taking the expectation value of the $n$-graviton particle projector,
\begin{align}
    \bar{P}_n &= \frac{1}{n!} \sum_{\sigma_1,...,\sigma_n = \pm} \int d \Phi(p_3) d \Phi(p_4) \int \prod_{i=1}^n d \Phi(k_i) |\langle k_1^{\sigma_1} ... k_n^{\sigma_n} p_3 p_4|S|p_1 p_2 \rangle|^2 .
    \label{eqn:probability}
\end{align}
As it is written, \eqref{eqn:probability} is formally divergent, as is known from the study of infrared divergences in quantum field theory (see \cite{bookItzykson}) because of the contribution of zero-energy gravitons. We will therefore work with a finite-resolution detector $\lambda > 0$, which implies that we will study only the probabilities of gravitons emitted with an energy $E_k > \lambda$. Correspondingly, we will replace
\begin{align}
\bar{P}_n \to \bar{P}_n^{\lambda}, \qquad \int \prod_{i=1}^n d \Phi(k_i) \to \int_{\lambda} \prod_{i=1}^n d \Phi(k_i),
\label{eqn:finite_reso}
\end{align}
As we will see later, we will not be interested in the single probability but in a particular infrared-safe combination of probabilities. Therefore $\lambda$ will be used only as an intermediate regulator, and in the end we will send $\lambda \to 0$.

We would like to scatter classically two massive point particles with classical momenta $m_A v_A$ and $m_B v_B$ with an impact parameter $b^{\mu}$. 
Since the main purpose of this paper is to take the classical limit from a quantum field theory calculation, we use the KMOC formalism \cite{Kosower:2018adc} and take instead as our incoming state 
\begin{align}
    | \psi_{\textrm{in}} \rangle := \int  d \Phi(p_1) d \Phi(p_2) e^{i b\cdot p_1/\hbar} \psi_A(p_1)  \psi_B(p_2) |p_1 p_2 \rangle ,
    \label{eqn:psi_in}
\end{align}
where
\begin{align}
d \Phi(p_1) := \frac{1}{(2 \pi)^3} d^4 p_1 \delta(p_1^2 - m_A^2) \theta(p_1^0), \qquad d \Phi(p_2) := \frac{1}{(2 \pi)^3} d^4 p_2 \delta(p_2^2 - m_B^2) \theta(p_2^0).
\end{align} 
The wavefunctions $\psi_A(p_1),\psi_B(p_2)$ are defined as
\begin{align}
\psi_A(p_1) := \mathcal{N}_1 m_A^{-1} \exp \left[-\frac{p_1 \cdot v_A}{\hbar \ell_{c,A} / \ell_{w,A}^{2}}\right], \qquad \psi_B(p_2) := \mathcal{N}_2 m_B^{-1} \exp \left[-\frac{p_2 \cdot v_B}{\hbar \ell_{c,B} / \ell_{w,B}^{2}}\right],
\label{eqn:wavef_def}
\end{align}
where ${\mathcal N}_1, \mathcal{N}_2$ are normalization factors, $\ell_{c,j} = \hbar / m_j$ is the Compton wavelength, and $\ell_{w,j}$ is related to the intrinsic spread of the wavefunction for the $j$-th massive particle ($j=A,B$). We will also require the ``Goldilocks conditions'' 
\begin{align}
\ell_{c,j} \ll \ell_{w,j} \ll b \qquad \text{for} \quad  j=A,B,
\end{align}
which ensure that wavefunctions such as those in \eqref{eqn:wavef_def} effectively localize
 the massive particles on their classical trajectories as $\hbar \to 0$. We expand the $S$-matrix in terms of the scattering matrix $T$,
\begin{align}
S=1 + i T.
\end{align}
For the expectation value of the graviton projector operator, only the amplitudes with at least one graviton emitted are going to contribute. We can read off from \eqref{eqn:probability} the probability of emitting $n$ gravitons with energies $E_{k_i} > \lambda$,
\begin{align}
   P_n^{\lambda} &= \frac{1}{n!} \sum_{\sigma_1,...,\sigma_n = \pm}  \int d \Phi(r_1) d \Phi(r_2) \int_{\lambda} \prod_{i=1}^n d \Phi(k_i) \langle\psi_{\textrm{in}}|  T^{\dagger} |r_1 r_2  k_1^{\sigma_1} ... k_n^{\sigma_n} \rangle \langle r_1 r_2  k_1^{\sigma_1} ... k_n^{\sigma_n}  |  T |\psi_{\textrm{in}}\rangle .
   \label{eqn:probability1exp}
\end{align}
We now expand $| \psi_{\textrm{in}} \rangle$ from (\ref{eqn:psi_in}) in terms of the wavefunction $\psi_A(p_1)\psi_B(p_2)$ and $\langle \psi_{\textrm{in}}| $ in terms of $\psi_A(p_1')\psi_B(p_2')$, 
and we introduce the momentum transfers \cite{Kosower:2018adc},
\begin{align}
\label{eq:momentumtransfers}
q_j := p_j' - p_j, \qquad w_j := r_j - p_j, \qquad q:=q_1=-q_2.
\end{align}
Writing the momentum integrals explicitly and inserting the momentum constraints and amplitudes, we then arrive at
\begin{align}
   P_n^{\lambda} &=  \frac{1}{n!} \sum_{\sigma_1,...,\sigma_n = \pm} \int d \Phi(p_1) d \Phi(p_2) \int_{\lambda} \prod_{i=1}^n d \Phi(k_i) \int \frac{d^4 q}{(2 \pi)^4} \int \prod_{j=1,2}  d^4 w_{j}   \nonumber \\
   &\times \delta\left(2 p_{1} \cdot q+q^{2}\right) \delta\left(2 p_{2} \cdot q-q^{2}\right) \Theta\left(p_{1}^{0}+q^{0}\right) \Theta\left(p_{2}^{0}-q^{0}\right) e^{-i b \cdot q / \hbar} \delta^{(4)}\left(w_{1}+w_{2}+\sum_{i=1}^n k_i \right)  \nonumber \\
   &\times \psi_A(p_1) \psi^*_A(p_1 + q) \psi_B(p_2) \psi^*_B(p_2 - q) \prod_{j=1,2} \left[ \delta\left(2 p_{j} \cdot w_{j}+w_{j}^{2}\right) \Theta\left(p_{j}^{0}+w_{j}^{0}\right)\right]\nonumber \\
&\times \mathcal{A}_{n+4} \left(p_{1}, p_{2} \rightarrow p_{1}+w_{1}, p_{2}+w_{2}, k_1^{\sigma_1},...,k_n^{\sigma_n}\right)  \nonumber \\
&\hspace{120pt} \times \mathcal{A}^{*}_{n+4}\left(p_{1}+q, p_{2}-q \rightarrow p_{1}+w_{1}, p_{2}+w_{2}, k_1^{\sigma_1},...,k_n^{\sigma_n}\right).
   \label{eqn:probability2}
\end{align}
 We now conveniently define a set of symmetrized variables for the external momenta \cite{Parra-Martinez:2020dzs}
\begin{align}
\label{eq:symm-mom}
p_A^{\mu} := p_1^{\mu} + \frac{q^{\mu}}{2}, \qquad p_B^{\mu} := p_2^{\mu} - \frac{q^{\mu}}{2},
\end{align}
which has the nice property of enforcing exactly the condition $p_A \cdot q = p_B \cdot q = 0$. In terms of these new variables \cite{Aoude:2021oqj}, we have
\begin{align}
P_n^{\lambda} &=  \frac{1}{n!} \sum_{\sigma_1,...,\sigma_n = \pm} \Big\langle\hspace{-3pt}\Big\langle \int_{\lambda} \prod_{i=1}^n d \Phi(k_i) \int \frac{d^4 q}{(2 \pi)^4}  \delta\left(2 p_{A} \cdot q\right) \delta\left(2 p_{B} \cdot q\right) \Theta\left(p_{A}^{0}+\frac{q^{0}}{2}\right) \Theta\left(p_{B}^{0}-\frac{q^{0}}{2}\right) \nonumber \\
   &\times \int d^4 w_1 d^4 w_2  e^{-i b \cdot q / \hbar}  \delta^{(4)}\left(w_{1}+w_{2}+\sum_{i=1}^n k_i \right)   \nonumber \\
   &\times \delta\left(2 p_A \cdot w_1+w_1^{2} - q \cdot w_1\right) \Theta\left(p_A^{0}+w_1^{0}- \frac{q^0}{2}\right)  \delta\left(2 p_B \cdot w_2+w_2^{2} + q \cdot w_2\right) \Theta\left(p_B^{0}+w_2^{0}+ \frac{q^0}{2}\right) \nonumber \\
&\times \mathcal{A}_{n+4} \left(p_A - \frac{q}{2}, p_B + \frac{q}{2} \rightarrow p_A+w_{1}- \frac{q}{2}, p_B+w_{2}+ \frac{q}{2}, k_1^{\sigma_1},...,k_n^{\sigma_n}\right)  \nonumber \\
&\hspace{50pt} \times \mathcal{A}^{*}_{n+4}\left(p_A+\frac{q}{2}, p_B-\frac{q}{2} \rightarrow p_A+w_{1}- \frac{q}{2}, p_B+w_{2}+ \frac{q}{2}, k_1^{\sigma_1},...,k_n^{\sigma_n}\right) \Big\rangle\hspace{-3pt}\Big\rangle ,
   \label{eqn:probability2exp}
\end{align}
where we use the double bracket notation $\Big\langle\hspace{-3pt}\Big\langle \cdot \Big\rangle\hspace{-3pt}\Big\rangle $ introduced in \cite{Kosower:2018adc}, which contains the implicit phase space integral over $p_{A}$, $p_{B}$ and the appropriate wavefunctions
\begin{align}
 \Big\langle\hspace{-3pt}\Big\langle f\left(p_{A}, p_{B}, \ldots\right)\Big\rangle\hspace{-3pt}\Big\rangle \equiv \int d \Phi(p_{A}) d \Phi(p_{B})  |\psi_A(p_A)|^2 |\psi_B(p_B)|^2 f\left(p_A, p_B, \ldots\right) ,
\end{align}
where
\begin{align}
d \Phi(p_A) &:= \frac{1}{(2 \pi)^3} d^4 p_A \delta\left(p_A^2 - m_A^2 + q^2/4\right) \theta\left(p_A^0-q^0/2\right),\nonumber \\
d \Phi(p_B) &:= \frac{1}{(2 \pi)^3} d^4 p_B \delta\left(p_B^2 - m_B^2 + q^2/4\right) \theta\left(p_B^0+q^0/2\right).
\end{align} 
Note that \eqref{eqn:probability2} is expressed in terms of unitarity cuts involving $n$ gravitons and the two massive particles in the intermediate state. The same result can be obtained by applying the LSZ reduction with the appropriate KMOC wavefunctions from the in-in formalism, as shown in appendix \ref{sec:appendix_SK}.

\subsection{Mean, variance and factorial moments of the graviton particle distribution}

In classical physics, we are interested in knowing whether the final graviton particle distribution is exactly Poissonian or super-Poissonian (the most general case). We refer the reader to appendix \ref{sec:Poissonian} for a brief review of the two cases. Poissonian statistics are known to be equivalent to having a single coherent state representing the quantum state for the classical radiation field. Here we give a short argument \cite{Cristofoli:2021vyo} for why we expect a single coherent state,  based on the fact that the we expect the incoming state to be a pure state in the classical limit and on the unitarity of the S-matrix. The work of Glauber in 1963 \cite{Glauber:1963tx,Glauber:1963fi} shows that every quantum state of radiation (i.e.\ every density matrix) can be written as a superposition of coherent states,
\begin{align}
\hat{\rho}_{k,\text{out}} = \sum_{\sigma = \pm} \int d^2 \alpha^{\sigma}_k \,\,\mathcal{P}_\textrm{out}^{\sigma}(\alpha_k) \ket{\alpha^{\sigma}_k} \bra{\alpha^{\sigma}_k},  \qquad 
d^2 \alpha^{\sigma}_k := \frac{d \Re(\alpha^{\sigma}_k) d \Im(\alpha^{\sigma}_k)}{\pi},
\label{eq:Glauber-SudarshanQM}
\end{align}
where $\mathcal{P}_\textrm{out}^{\sigma}(\alpha_k)$ is a well-defined probability density ($\mathcal{P}_\textrm{out}^{\sigma}(\alpha_k) \geq 0$) in the coherent state space in the classical limit, and $\ket{\alpha^{\sigma}_k}$ represents a coherent state of a graviton excitation (``harmonic oscillator'') of momentum $k$ and definite helicity $\sigma$, which we can write generically as
\begin{align}
\ket{\alpha^{\sigma}_k} :=\exp\biggl[\alpha_k a^{\dagger}_{\sigma}(k) - \alpha^*_k a_{\sigma}(k) \biggr] \ket{0} ,
\end{align}
where $a^{\dagger}_{\sigma}(k)$ and $a_{\sigma}(k)$ are the creation and annihilation operators of a graviton of helicity $\sigma$.
This representation is known as the Glauber-Sudarshan P-representation \cite{Sudarshan:1963ts,Glauber:1963tx}, and it is widely used in the quantum optics literature. In quantum field theory, we need to consider an infinite superposition of harmonic oscillators for all momenta $k \in \mathbb{R}^{1,3}$, and therefore we will promote \eqref{eq:Glauber-SudarshanQM} to\footnote{See also \cite{Cristofoli:2021jas} for a more rigorous approach by taking the large volume limit of a finite spacetime box, where momenta are quantized and we only need to consider a finite superposition of harmonic oscillators. We thank Donal O'Connell for emphasizing this point.}
\begin{align}
\hat{\rho}_{\text{radiation,out}} = \sum_{\sigma = \pm} \int \mathcal{D}^2 \alpha^{\sigma} \, \mathcal{P}_\textrm{out}^{\sigma}(\alpha) \ket{\alpha^{\sigma}} \bra{\alpha^{\sigma}}  ,
\label{eq:Glauber-Sudarshan}
\end{align}
where now 
\begin{align}
\ket{\alpha^{\sigma}} =\exp\biggl[\int d \Phi(k)  (\alpha(k) a^{\dagger}_{\sigma}(k) - \alpha^*(k) a_{\sigma}(k)) \biggr] \ket{0} .
\end{align}
Since we are dealing with scattering boundary conditions and our incoming KMOC state $\ket{\psi_{\text{in}}}$ is a pure state, the unitarity of the S-matrix $S S^{\dagger} = 1$ implies that $\ket{\psi_{\text{in}}}$ is mapped to outgoing pure states. Therefore,  the outgoing radiation state must be a superposition of pure states,
\begin{align}
\mathcal{P}^{\sigma}_{\textrm{out}}(\alpha) = \sum_{j=1}^{\infty} c^{\sigma}_{j,\textrm{out}} \delta^2(\alpha^{\sigma} - \alpha_j^{\sigma}) .
\end{align}
But thanks to a crucial theorem of Hillery \cite{HILLERY1985409}, we know that every such superposition of pure states is trivial in the classical limit $\hbar \to 0$,
\begin{align}
\mathcal{P}^{\sigma}_{\textrm{out},\star}(\alpha) = \delta^2(\alpha^{\sigma} - \alpha^{\sigma}_{\star}).
\end{align}
We therefore expect, on general grounds, to be able to describe the final radiation state for a scattering process involving point particles with a single coherent state.

From the pure amplitude perspective, the same question is hard to answer unless we work strictly in the soft approximation \cite{Ware:2013zja,Addazi:2019mjh,Gonzo:2020xza}. But in general, we can address this question perturbatively by studying the mean, the variance and the factorial moments of the particle distribution. A similar approach has been taken by F.\ Gelis and R.\ Venugopalan \cite{Gelis:2006yv,Gelis:2006cr,Gelis:2015kya} in the standard in-in formalism, which we try to specialize here from a fully on-shell perspective and in the classical limit. 

\begin{figure}[h]
\centering
\includegraphics[scale=0.65]{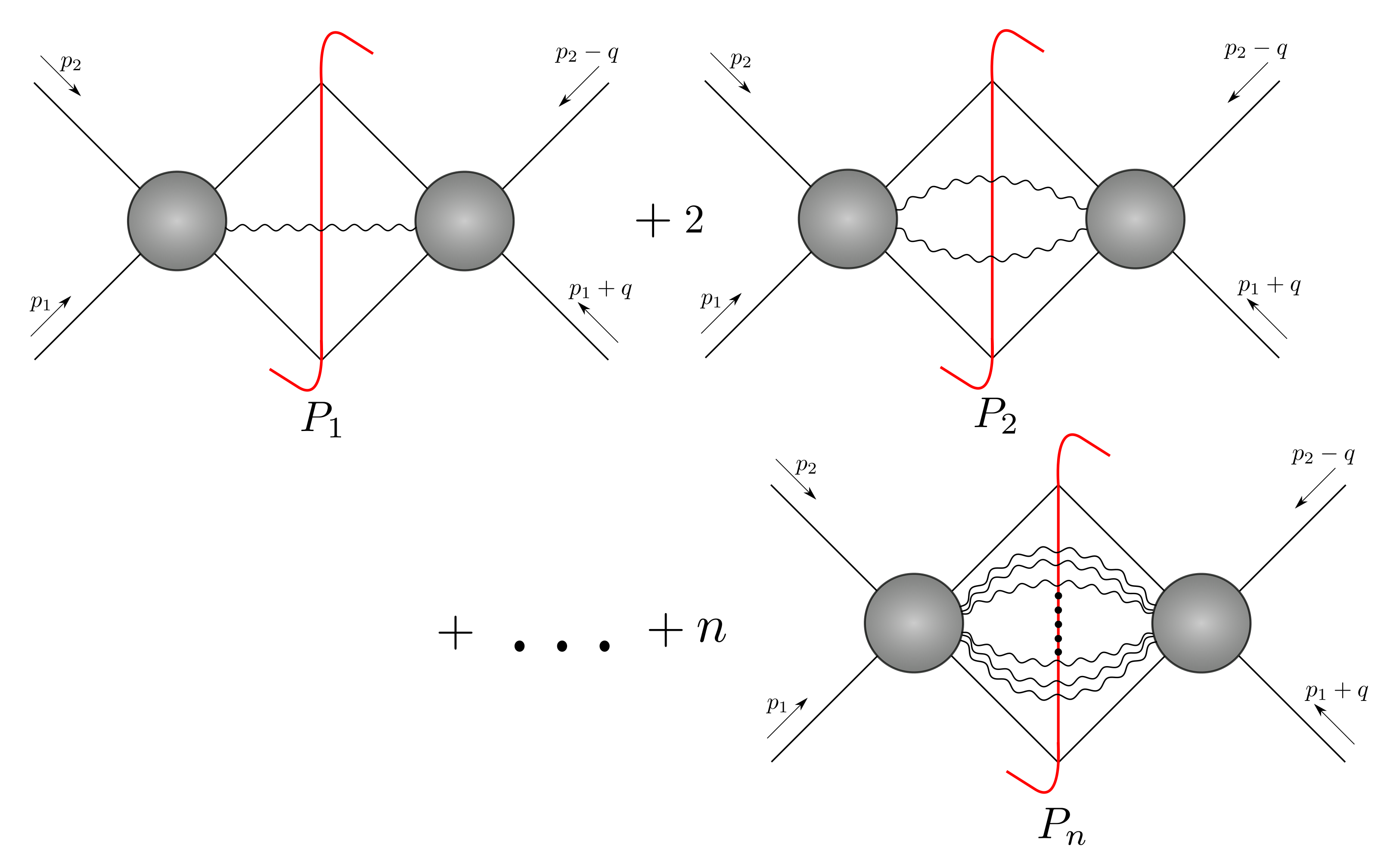}
\caption{Diagrammatic representation of the on-shell amplitude contribution to the graviton number operator expectation value.}
\label{fig:number_oper}
\end{figure}

The graviton number operator is defined as
\begin{align}
     \hat{N} = \sum_{\sigma=\pm} \int d \Phi(k) \, a^{\dagger}_{\sigma}(k) a_{\sigma}(k) .
\end{align}
  Having defined 
\begin{align}
     | \psi_{\textrm{out}} \rangle := S | \psi_{\textrm{in}} \rangle ,
\end{align}
the expectation value of the number operator in the final state gives the mean of the distribution, which can be expressed in terms of unitarity cuts in a similar fashion to the derivation of equation~\bref{eqn:probability2}, as depicted in Figure~\ref{fig:number_oper}. The mean of the distribution is defined as
\begin{align}
    \mu_{\textrm{out}}^{\lambda} &:= \langle \psi_{\textrm{out}}| \hat{N} | \psi_{\textrm{out}} \rangle \nonumber \\
    &= \int d \Phi(r_1) d \Phi(r_2) \sum_{n_X} \int_{\lambda} d \Phi(X) n_X  \langle\psi_{\textrm{in}}|  T^{\dagger} |r_1 r_2 X \rangle \langle r_1 r_2 X |  T |\psi_{\textrm{in}}\rangle \nonumber \\
    &= \sum_{n=0}^{\infty} n \, P_n^{\lambda} ,
    \label{eqn:N_exp}
\end{align}
where $\ket{r_1 r_2 X}$ denotes the state with $n_X$ gravitons and two massive particles of momenta $r_1$ and $r_2$, and $\int_{\lambda} d \Phi(X)$ stands for the phase space integration for the gravitons.

We define the variance of the distribution as
\begin{align}
\Sigma_{\textrm{out}}^{\lambda} &:= \langle\psi_{\textrm{out}}| (\hat{N})^2 |\psi_{\textrm{out}}\rangle - \left(\langle\psi_{\textrm{out}}| \hat{N} |\psi_{\textrm{out}}\rangle\right)^2 = \sum_{n=0}^{\infty} n^2 \, P_n^{\lambda} - \left(\sum_{n=0}^{\infty} n \, P_n^{\lambda}\right)^2 .
\end{align}
If the variance is equal to the mean, i.e.\ if
\begin{align}
\Sigma_{\textrm{out}}^{\lambda} \stackrel{?}{=} \mu_{\textrm{out}}^{\lambda},
\end{align}
then the distribution is consistent with a Poissonian distribution. This means that the deviation from the Poissonian distribution,
\begin{align}
\Delta_{\textrm{out}} :&= \Sigma_{\textrm{out}}^{\lambda} - \mu_{\textrm{out}}^{\lambda}  \nonumber \\
&= \sum_{n=0}^{\infty} (n^2 - n) \, P_n^{\lambda} - \left(\sum_{n=0}^{\infty} n \, P_n^{\lambda}\right)^2,
\label{eqn:Delta_def}
\end{align}
characterizes the deviation from the coherent state description.

We claim here that the difference between the mean $\mu^{\lambda}$ and the variance $\Sigma^{\lambda}$ is an infrared-safe quantity in perturbative quantum gravity. While the probability of emission of $n$ gravitons is generally ill-defined because of infrared divergences, there is a non-trivial cancellation which happens for $\Delta_{\textrm{out}}$. Indeed, the contribution of zero-energy gravitons to the final state, which give rise to the infrared divergent contributions, is known to be exactly represented by a coherent state. This can be proved either from a Faddeev-Kulish approach \cite{Ware:2013zja,Gonzo:2020xza} or from a path integral perspective \cite{Laenen:2008gt,White:2011yy,Cristofoli:2021jas}. Let us denote the mean and the variance of this coherent state for zero-energy gravitons by $\mu_{\textrm{out}}^{E_k \sim 0}$ and $\Sigma_{\textrm{out}}^{E_k \sim 0}$ respectively. In  Appendix \ref{sec:Poissonian}, we show that for such a coherent state of soft gravitons we have\footnote{It is not necessary to specify $\alpha^{\sigma}_{E_k \sim 0}(k)$ for the argument to work. The interested reader can find additional details in \cite{Ware:2013zja}.} 
\begin{align}
\Sigma_{\textrm{out}}^{E_k \sim 0} &= \mu_{\textrm{out}}^{E_k \sim 0} = \sum_{\sigma = \pm} \int_{E_k \sim 0} d \Phi(k) |\alpha^{\sigma}_{E_k \sim 0}(k)|^2, \nonumber \\
\Delta_{\textrm{out}}^{E_k \sim 0} &= \Sigma_{\textrm{out}}^{E_k \sim 0} - \mu_{\textrm{out}}^{E_k \sim 0} = 0.
\end{align}
This is the reason why the cutoff $\lambda$ was removed in \eqref{eqn:Delta_def}.\footnote{This argument does not apply directly to non-abelian theories because of the presence of collinear divergences, which for perturbative gravity are known to cancel exactly \cite{Akhoury:2011kq}. It would be interesting to develop this idea further, along the lines of \cite{delaCruz:2020bbn,delaCruz:2021gjp}.}

We can easily check by induction on the number of loops and legs that\footnote{To avoid cluttering the notation, we keep the $\lambda$ dependence implicit in $P_n^{(L_1,L_2)} \equiv P_n^{(L_1,L_2),\lambda}$.}
\begin{align}
P_n^{\lambda} = \sum_{L_1,L_2 = 0}^{\infty} G^{2 + n + L_1 + L_2} P_n^{(L_1,L_2)}, 
\end{align}
where we have  explicitly extracted the scaling in the gravitational coupling $G$ of the product of an $L_1$-loop amplitude with an $L_2$-loop amplitude with $n$ gravitons.
The lowest order contribution to $\Delta_{\textrm{out}}$ is of order $\mathcal{O}(G^4)$, which corresponds to
\begin{align}
\Delta_{\textrm{out}} \Big|_{\mathcal{O}(G^4)} = 2 G^4 P_2^{(0,0)}.
\label{eqn:leading_Poissonian_dev}
\end{align}
This leading term is the unitarity cut involving the 6-pt tree amplitude $\mathcal{A}^{(0)}_6(\phi_A \phi_B \to \phi_A \phi_B h_1 h_2)$ and its conjugate $\mathcal{A}^{(0)*}_6(\phi_A \phi_B \to \phi_A \phi_B h_1 h_2)$. It is important also to understand the higher order terms in $\Delta_{\textrm{out}}$, since they will give non-trivial amplitude relations if we assume coherence at all orders. From the definition \eqref{eqn:Delta_def}, we have
\begin{equation}
   \boxed{
   \begin{aligned}
      \Delta_{\textrm{out}} &= \sum_{L_1,L_2 = 0}^{\infty} \sum_{n = 2}^{\infty} G^{2 + n + L_1 + L_2} (n^2 - n) P_n^{(L_1,L_2)} \\
&\qquad\qquad - \sum_{L_1,L_2,L_1',L_2' = 0}^{\infty} \sum_{n,m = 1}^{\infty} n m \,G^{4 + n + m + L_1 + L_2 + L_1' + L_2'} P_n^{(L_1,L_2)} P_m^{(L_1',L_2')}.
   \end{aligned}
   }
\label{eqn:main_result_Delta}
\end{equation}
Let us examine the first several terms appearing explicitly in the expansion of \eqref{eqn:main_result_Delta},
\begin{align}
&\Delta_{\textrm{out}}  = 2 G^4 P_2^{(0,0)} + 6 G^5 P_3^{(0,0)} + 12 G^6 P_4^{(0,0)} + 20 G^7 P_5^{(0,0)}  \nonumber \\
& + G^5 (2 P_2^{(1,0)} + 2 P_2^{(0,1)} ) + G^6 ( 2 P_2^{(0,2)} + 2 P_2^{(2,0)} + 6 P_3^{(1,0)} + 6 P_3^{(0,1)} ) \nonumber \\
& + G^7 (2 P_2^{(3,0)} + 2 P_2^{(0,3)} + 6 P_3^{(2,0)} + 6 P_3^{(0,2)} + 6 P_3^{(1,1)} + 12 P_4^{(1,0)} + 12 P_4^{(0,1)} - 4 P_1^{(0,0)} P_2^{(0,0)}) \nonumber \\
& + \Big[G^6 (2 P_2^{(1,1)} - (P_1^{(0,0)})^2) + G^7 (2 P_2^{(1,2)} + 2 P_2^{(2,1)} - 2 P_1^{(0,1)} P_1^{(0,0)} - 2 P_1^{(1,0)} P_1^{(0,0)})\Big],
\label{eqn:Delta_expansion}
\end{align}
where we have organized each different line according to the expected behavior of the terms in the classical limit. We expect that the first three lines of \eqref{eqn:Delta_expansion} are related to ``quantum'' contributions and are therefore irrelevant in the classical limit. The last line of \eqref{eqn:Delta_expansion}, instead, contains a combination of unitarity cuts which will give non-trivial quadratic relations between ``classical'' loop amplitudes with a higher number of emitted gravitons of the form $P_n^{(L_1,L_2)}$ with $n \geq 2$ and $L_1 + L_2 \geq 1$, and 5-point amplitude contributions involving $P_1^{(L_1,L_2)}$. We will discuss this interpretation in more detail in section \ref{sec:classical_limit}, where we will also emphasize the relevance of the 5-pt amplitude for the calculation of classical radiative observables.

It is important to consider also higher moments of the statistical distribution for the graviton number production. We can define a generating functional 
\begin{align}
F(x) = \sum_{n=0}^{\infty} P_n^{\lambda} e^{n x} ,
\end{align}
from which all higher moments can be derived, 
\begin{align}
\langle \psi_{\textrm{out}}| \hat{N}^m | \psi_{\textrm{out}} \rangle &= \sum_{n=0}^{\infty} n^m \, P_n^{\lambda} = \frac{d^m F(x)}{d x^m} \Big|_{x = 0} .
    \label{eqn:N_moments}
\end{align}
Therefore, the knowledge of all graviton emission probabilities $P_n^{\lambda}$ is enough to completely determine the distribution of the particles above the energy cutoff. In practice, we can rely on perturbation theory and therefore computing the first few moments is enough to accurately determine the particle distribution. We can also defined connected moments (or ``cumulants''), like the variance and its higher order generalizations. Having defined a generating functional 
\begin{align}
G(x):= \log(F(x)) ,
\end{align}
for a Poissonian distribution we would expect, given a certain waveshape $\alpha^{\sigma}(k)$, that
\begin{align}
\Sigma^{(m),\lambda}_{\text{Poisson}} &:= \frac{d^m G^{\text{Poisson}}(x)}{d x^m} \Big|_{x = 0} \sim \sum_{\sigma = \pm} \int_{\lambda} d \Phi(k) |\alpha^{\sigma}(k)|^2 \qquad \text{for all~}  m >0.
\end{align}
because all the cumulants should be equal.
In particular, the variance is a special case for $m=2$, i.e.\ $\Sigma^{(2),\lambda} = \Sigma^{\lambda}$. 

For our purposes it is more convenient to consider factorial moments $\Gamma^{(m)}$, which correspond to a linear combination of the connected moments discussed above. We define the factorial moments
\begin{align}
\Gamma^{(m),\lambda}_{\textrm{out}} &:= \langle \psi_{\textrm{out}}| \prod_{j=1}^m (\hat{N} - j + 1) | \psi_{\textrm{out}} \rangle \nonumber \\
&= \langle \psi_{\textrm{out}}| \hat{N} (\hat{N} - 1) \dots (\hat{N} - m + 1) | \psi_{\textrm{out}} \rangle \,\,.
\label{eq:factorial-moments}
\end{align}
For a Poissonian distribution it is possible to prove that 
\begin{align}
\Gamma^{(m)}_{\textrm{out},\text{Poisson}} = (\mu_{\textrm{out},\text{Poisson}}^{\lambda})^m
\end{align}
and therefore we can also consider in perturbation theory other infrared-safe combinations of probabilities like
\begin{align}
\Delta^{(m)}_{\textrm{out}} := \Gamma_{\textrm{out}}^{(m),\lambda} - (\mu_{\textrm{out}}^{\lambda})^m,
\label{eq:Delta_higher}
\end{align}
where for $m=2$ one can check that we recover the difference between the mean and the variance in \eqref{eqn:main_result_Delta}. By expanding \eqref{eq:Delta_higher} we get immediately
\begin{align}
   \boxed{
   \begin{aligned}
      \Delta^{(m)}_{\textrm{out}} =& \sum_{n=0}^{\infty} \sum_{L_{1}, L_{2}=0}^{\infty} G^{2+n+L_{1}+L_{2}} \frac{n!}{(n-m)!} P_{n}^{\left(L_{1}, L_{2}\right)} \\
&- \sum_{n_{1}, \ldots, n_{m}=1}^{\infty} \sum_{L_{1}^{(1)}, \ldots, L_{1}^{(m)}=0}^{\infty} \sum_{L_{2}^{(1)}, \ldots, L_{2}^{(m)}=0 }^{\infty} G^{2 m+\sum_{k=1}^{m}\left[n_{k}+L_{1}^{(k)}+L_{2}^{(k)}\right]} \prod_{j=1}^{m}\left[n_{j} P_{n_{j}}^{\left(L_{1}^{(j)}, L_{2}^{(j)}\right)}\right].
   \end{aligned}
   }
   \label{eqn:main_result_Deltam}
\end{align}
It is interesting to consider the first terms in this expansion of $\Delta^{(3)}_{\textrm{out}}$,
\begin{align}
\Delta^{(3)}_{\textrm{out}}  =& 6 G^5 P_3^{(0,0)} + 24 G^6 P_4^{(0,0)} + 60 G^7 P_5^{(0,0)} \nonumber \\
& + G^6 (6 P_3^{(1,0)} + 6 P_3^{(0,1)} ) + G^7 (6 P_3^{(0,2)} + 6 P_3^{(2,0)} + 6 P_3^{(1,1)} + 24 P_4^{(1,0)} + 24 P_4^{(0,1)} ) \, ,
\label{eqn:Delta3_expansion}
\end{align}
and of $\Delta^{(4)}_{\textrm{out}}$,
\begin{align}
\Delta^{(4)}_{\textrm{out}}  =& 24 G^6 P_4^{(0,0)} + 120 G^7 P_5^{(0,0)} \nonumber \\
& + G^7 (24 P_4^{(1,0)} + 24 P_4^{(0,1)} )\, ,
\label{eqn:Delta4_expansion}
\end{align}
where we have organized the terms similarly to what was done in \eqref{eqn:Delta_expansion}.
We will explore the deep consequences of assuming coherence at all orders, i.e. $\Delta^{(m)}_{\textrm{out}} = 0$, in section \ref{sec:classical_limit}. In \cite{Cristofoli:2021jas}, it is shown how coherence properties are linked to the factorization of radiative observables in the KMOC formalism.\footnote{Similar statements about the classical factorization have been made in \cite{Gonzo:2020xza} for infrared divergences and in \cite{Cristofoli:2021vyo} for the classical expansion.} In classical physics, we expect only the 1-point function to play a role for any observable of interest. Such an observable is essentially uniquely determined by the classical equations of motion and the retarded boundary conditions at $t \to -\infty$: all two-point and higher-point functions then have to factorize as $\hbar \to 0$. There the following relation was established,
\begin{equation*}
\begin{aligned}[t]
\text{Poissonian distribution} \\
\text{in Fock space}\hspace{20pt}\\
\end{aligned}
\qquad\Longleftrightarrow\qquad
\begin{aligned}[t]
\text{Zero-variance property}\hspace{50pt}\\
\text{in the Glauber-Sudarshan coherent state basis}\\
\end{aligned}
\end{equation*}
which implies that Poissonian distributions in the number operator basis correspond to a degenerate distribution ($\propto \delta^2(\alpha^{\sigma} - \alpha^{\sigma}_{\star})$) in the Glauber-Sudarshan space.

\section{Tree amplitudes from Feynman diagrams}
\label{sec:Feynman}

In this section, we extend the parametrization of the pure lagrangian used by Cheung and Remmen \cite{Cheung:2017kzx} to the case of real scalar fields minimally coupled with gravity. This will make use of an auxiliary field, the connection, whose job is to effectively resum higher order graviton pure contact vertices in the same spirit as the first order Palatini formulation developed by Deser \cite{Deser:1969wk,Deser:1987uk}. We can then compute in a straightforward way all the tree level amplitudes we need for this work.

Let us consider the lagrangian of two real scalars minimally coupled with gravity in $D=4$ dimensions,
\begin{align}
S &:= S_{GR} + S_{matter}, \nonumber \\
S_{GR} &:= \frac{1}{16 \pi G} \int d^4 x \, \left[\partial_a \sigma_{c e} \partial_b \sigma^{d e} \left( \frac{1}{4} \sigma^{a b} \delta^c_d - \frac{1}{2} \sigma^{c d} \delta^a_d \right) + \frac{1}{2} \sigma^{a b} \omega_a \omega_b \right], \nonumber \\
S_{matter} &:= -\sum_{j=A,B} \int d^4 x \, \left[\frac{1}{2} \sigma^{a b} \partial_a \phi_j \partial_b \phi_j + \frac{1}{2} \sqrt{-\det(\sigma^{-1})} m_j^2 \phi_j^2\right] ,
\label{eqn:lagrangian_GRmatter}
\end{align}
where we work in the mostly-plus signature and have used the following conventions:
\begin{align}
\sigma_{a b}&:=\frac{1}{\sqrt{-g}} g_{a b}, \quad \sigma^{a b}=\sqrt{-g} g^{a b}, \quad \det(g)=\det(\sigma^{-1}), \nonumber \\
\omega_{a} &:=\partial_{a} \log \sqrt{-g}=\frac{1}{2} \sigma_{b c} \partial_{a} \sigma^{b c} .
\end{align}
We introduce the auxiliary field $A^a_{b c}$, which allows us to rewrite the pure gravity lagrangian as 
\begin{align}
S_{GR} &= \frac{1}{16 \pi G} \int d^4 x \, \left[-\left(A_{b c}^{a} A_{a d}^{b}-\frac{1}{3} A_{a c}^{a} A_{b d}^{b}\right) \sigma^{c d}+A_{b c}^{a} \partial_{a} \sigma^{b c}\right] .
\end{align}
Before setting up the perturbation theory in the new variables, it is useful to unmix the graviton and auxiliary field by doing the shift
\begin{align}
A_{b c}^{a} \rightarrow A_{b c}^{a}-\frac{1}{2}\left(\partial_{b} h_{c}^{a}+\partial_{c} h_{b}^{a}-\partial^{a} h_{b c}+\frac{1}{2} \eta_{b c} \partial^{a} h_{d}^{d}\right)
\end{align}
and adding the gauge fixing term
\begin{align}
\mathcal{L}_{\mathrm{GF}}=-\frac{1}{2} \partial_{a} h^{a c} \partial^{b} h_{b c}=-\frac{1}{2} \eta_{c d} \partial_{a}\left(\sqrt{-g} g^{a c}\right) \partial_{b}\left(\sqrt{-g} g^{b d}\right) .
\end{align}
Using
\begin{align}
\sigma^{a b}=\eta^{a b} - \kappa h^{a b}
\end{align}
with $\kappa = \sqrt{32 \pi G}$, and the expansion
\begin{align}
\sqrt{-\det(\sigma^{-1})} &= \exp\left[\frac{1}{2} \Tr\log\left(1-\kappa \eta h^{-1}\right)\right] \nonumber \\
&= 1- \frac{\kappa}{2} \Tr\left(\eta h^{-1}\right) - \frac{\kappa^2}{4} \Tr\left(\eta h^{-1}\right)^2 - \frac{\kappa^3}{8} \Tr\left(\eta h^{-1}\right)^3 + \frac{\kappa^2}{8} \Tr^2\left(\eta h^{-1}\right) \nonumber \\
&\hspace{10pt}+ \frac{\kappa^3}{6} \Tr\left(\eta h^{-1}\right) \Tr\left(\eta h^{-1}\right)^2 + \mathcal{O}(h^4)\nonumber \\
&= 1 - \frac{\kappa}{2} h^{a}_{a} - \frac{\kappa^2}{4} h^{a b} h_{a b} - \frac{\kappa^3}{8} h^{b c} h_{a b} h_{c}^{a} + \frac{\kappa^2}{8} (h_{a}^{a})^2 + \frac{\kappa^3}{6} h_{a}^{a} \left(h_{b c} h^{b c}\right) + \mathcal{O}(h^4) ,
\end{align}
we get explicitly up to $\mathcal{O}(h^4)$ a lagrangian of the form
\begin{align}
\mathcal{L}=\mathcal{L}_{\mathrm{GR}}+\mathcal{L}_{\mathrm{matter}}+\mathcal{L}_{\mathrm{GF}}&=\mathcal{L}_{h h}+\mathcal{L}_{A A}+\mathcal{L}_{h h h} \nonumber \\
&+\mathcal{L}_{h h A}+\mathcal{L}_{h A A} +\mathcal{L}_{\phi \phi} +\mathcal{L}_{h \phi \phi} +\mathcal{L}_{h h \phi \phi} +\mathcal{L}_{h h h\phi \phi}.
\end{align}
The quadratic terms in the lagrangian are given by
\begin{align}
\mathcal{L}_{h h} &:= \frac{1}{2}\left(h_{a b} \square h^{a b}-\frac{1}{2} h^e_e \square h^f_f\right), \nonumber \\
\mathcal{L}_{A A} &:=- 2 \left(A_{b c}^{a} A_{a d}^{b}-\frac{1}{3} A_{a c}^{a} A_{b d}^{b}\right) \eta^{c d}, \nonumber \\
\mathcal{L}_{\phi \phi} &:= -\sum_{j=A,B} \left[\frac{1}{2} \partial^a \phi_j \partial_a \phi_j + m_j^2 \phi_j^2 \right], 
\end{align}
and the interaction terms are\footnote{We define the symmetrized (resp.\ antisymmetrized) product for any tensorial expression $T$ as $T_{(a b)} = T_{a b} + T_{b a}$ (resp.\ $T_{[a b]} = T_{a b} - T_{b a}$).}
\begin{align}
\mathcal{L}_{h h h} &:=\kappa \frac{1}{2} h^{a b}\left[\partial_{a} h_{c d} \partial_{b} h^{c d}+2 \partial_{[c} h_{d ] b} \partial^{d} h_{a}^{c}+\frac{1}{2}\left(2 \partial_{c} h_{a b} \partial^{c} h^e_e-\partial_{a} h^e_e \partial_{b} h^f_f\right)\right], \nonumber \\
\mathcal{L}_{h h A} &:=2 \kappa  h^{a b}\left[A_{a d}^{c}\left(\partial^{d} h_{b c}-\partial_{(b} h_{c)}^{d}\right)-\frac{1}{2}\left(\eta_{a d} A_{b c}^{d} \partial^{c} h^e_e-A_{c a}^{c} \partial_{b} h^e_e\right)\right], \nonumber \\
\mathcal{L}_{h A A} &:=2 \kappa h^{a b}\left(A_{a d}^{c} A_{b c}^{d}-\frac{1}{3} A_{a c}^{c} A_{b d}^{d}\right), \nonumber \\
\mathcal{L}_{h \phi \phi} &:= \frac{\kappa}{2}\sum_{j=A,B} \left[h^{a b} \partial_a \phi_j \partial_b \phi_j + \frac{1}{2} h^{a}_a m_j^2 \phi_j^2 \right], \nonumber \\
\mathcal{L}_{h h \phi \phi} &:= \frac{\kappa^2}{8}\sum_{j=A,B} \left[h^{a b} h_{a b} - \frac{1}{2} (h_{a}^{a})^2\right]   m_j^2 \phi_j^2 ,\nonumber \\
\mathcal{L}_{h h h\phi \phi} &:= \frac{\kappa^3}{16}\sum_{j=A,B} \left[\frac{4}{3}  h^{b c} h_{a b} h_{c}^{a} - h_{a}^{a} \left(h_{b c} h^{b c}\right)\right] m_j^2 \phi_j^2 .
\end{align}
In the massless limit $m_A, m_B \to 0$, the interaction terms become purely trivalent. In that case, it is possible to set up the standard Berends-Giele recursion relations. But even with the mass terms, the final expressions are more compact than in the standard perturbative expansion of gravity: the gravity pure self-interactions are nicely resummed by the auxiliary field, which makes it possible to avoid the cumbersome expressions for higher point vertices (at least at tree level, where ghosts are absent). The Feynman rules for the propagators are then 
\begin{align}
\hspace{-30pt}(\Delta^{h h})_{a b c d}(p) &=-\frac{i}{2 p^{2}}\left(\eta_{a c} \eta_{b d}+\eta_{a d} \eta_{b c}-\eta_{a b} \eta_{c d}\right), \nonumber \\
\hspace{-30pt}(\Delta^{A A})_{b c e f}^{a d}(p) &=-\frac{i}{4}\left[\frac{1}{2} \delta_{(b}^{d} \eta_{c)(e} \delta_{f)}^{a}+\eta^{a d}\left(\frac{1}{2} \eta_{b c} \eta_{e f}-\frac{1}{2} \eta_{b(e} \eta_{f) c}\right)\right], \nonumber \\
\hspace{-30pt}(\Delta^{\phi_j \phi_j})(p) &=  -\frac{i}{p^{2} + m_j^2}\qquad \textrm{for~} j=A,B ,
\end{align}
and the rules for the interaction vertices are
{\allowdisplaybreaks
\begin{align}
\left\langle h_{a b} h_{c d} h_{e f}\right\rangle\left(p_{1}, p_{2}, p_{3}\right) &= i\frac{\kappa}{2}\Bigg\{\left[\frac{1}{2}\left(\eta_{a(c} \eta_{d)(e} \eta_{f) b}+\eta_{b(c} \eta_{d)(e} \eta_{f) a}\right)\left(p_{1} \cdot p_{2}\right) \right.\nonumber \\
&\left. \hspace{15pt} -\frac{1}{2}\left(\eta_{a b} \eta_{c(e} \eta_{f) d}+\eta_{c d} \eta_{a(e} \eta_{f) b}\right)\left(p_{1} \cdot p_{2}\right) \right.\nonumber \\
 &\hspace{15pt}\left.+\left(\frac{1}{2} \eta_{a b} \eta_{c d}-\frac{1}{2} \eta_{a(c} \eta_{d) b}\right) p_{1(e} p_{2 f)}-\frac{1}{2} p_{2(a} \eta_{b)(e} \eta_{f)(d} p_{1 c)}\right] \nonumber \\
&\hspace{15pt} +\left[\begin{array}{l}
p_{2} \leftrightarrow p_{3} \\
c d \leftrightarrow e f
\end{array}\right]+\left[\begin{array}{l}
p_{1} \leftrightarrow p_{3} \\
a b \leftrightarrow e f
\end{array}\right]\Bigg\}, \nonumber \\
\hspace{-30pt}\left\langle h_{a b} A_{d e}^{c} A_{g h}^{f}\right\rangle\left(p_{1}, p_{2}, p_{3}\right) &=i\frac{\kappa}{2}\left(\delta_{(g}^{c} \eta_{h)(a} \eta_{b)(d} \delta_{e)}^{f}-\frac{1}{3} \delta_{(g}^{f} \eta_{h)(a} \eta_{b)(d} \delta_{e)}^{c}\right), \nonumber \\
\hspace{-30pt}\left\langle h_{a b} h_{c d} A_{f g}^{e}\right\rangle\left(p_{1}, p_{2}, p_{3}\right)&=\frac{\kappa}{2}\Bigg\{\left[\frac{1}{2} \delta_{(a}^{e}\left(\eta_{b)(f} \eta_{g)(c} p_{1 d)}-\eta_{b)(c} \eta_{d)(f} p_{1 g)}\right)\right. \nonumber \\
&\hspace{15pt}\left.+\frac{1}{2} \eta_{a b}\left(p_{1(f} \eta_{g)(c} \delta_{d)}^{e}-p_{1(c} \eta_{d)(f} \delta_{g)}^{e}\right)\right] +\left[\begin{array}{c}
p_{1} \leftrightarrow p_{2} \nonumber\\
a b \leftrightarrow c d
\end{array}\right]\Bigg\}  \nonumber \\
&\qquad -\frac{\kappa}{4} p_{3}^{e}\left(\eta_{f(a} \eta_{b)(c} \eta_{d)g}+\eta_{g(a} \eta_{b)(c} \eta_{d) f}\right), \nonumber \\
\hspace{-30pt}\left\langle h_{a b} \phi_j \phi_j\right\rangle\left(p_{1}, p_{2}, p_{3}\right)  &= -i\frac{\kappa}{2} \left(p_{2 (a} p_{3 b)}-m_j^2 \eta_{a b}\right) \qquad \textrm{for~} j=A,B, \nonumber \\
\hspace{-30pt}\left\langle h_{a b} h_{c d} \phi_j \phi_j\right\rangle\left(p_{1}, p_{2}, p_{3},p_{4}\right)  &= i \frac{\kappa^2}{4} m^2_j \left(\eta_{a c} \eta_{b d} + \eta_{a d} \eta_{b c} - \eta_{a b} \eta_{c d}\right) \qquad \textrm{for~}j=A,B , \nonumber \\
\hspace{-30pt}\left\langle h_{a b} h_{c d} h_{e f} \phi_j \phi_j\right\rangle\left(p_{1}, p_{2}, p_{3},p_{4},p_{5}\right)  &= i\frac{\kappa^3}{8} m^2_j \left(\eta_{f a} \eta_{b (c} \eta_{d) e}+\eta_{d (a} \eta_{e) c} \eta_{f b} +\eta_{a (d} \eta_{e) b} \eta_{f c}+\eta_{e (a} \eta_{b) c} \eta_{f d} \right. \nonumber \\
&\hspace{40pt}\left. -[\eta_{a (c} \eta_{d) b} \eta_{e f}+\eta_{a (e} \eta_{f) b} \eta_{c d} +\eta_{a b} \eta_{e (c} \eta_{d) f} ]\right)\qquad \textrm{for~} j=A,B ,
\end{align}
}
where all momenta are chosen to be ingoing. At this point one can implement these Feynman rules in the xAct package \cite{xAct}, which we use extensively in the following calculations. 

For the purposes of simplifying computations, we adopt the following conventions for the momenta of our amplitude:
\begin{align}
	\mathcal{A}^{(0)}_{n+4}({\bf 1}^A,{\bf 2}^B,{\bf 3}^A,{\bf 4}^B,5^{\sigma_1},\dots,(n+4)^{\sigma_{n}})
	&\equiv \mathcal{A}^{(0)}_{n+4} \left( p_1,p_2 \rightarrow -p_3,-p_4, -p_5^{\sigma_1},\dots,-p_{n+4}^{\sigma_{n}}\right) \nonumber \\
	&\equiv \mathcal{A}^{(0)}_{n+4} \left( p_1,p_2 \rightarrow -p_3,-p_4, -k_1^{\sigma_1},\dots,-k_n^{\sigma_{n}}\right),
	\label{eq:conv}
\end{align}
and we define the momentum invariants $s_{ij}=-(p_i+p_j)^2$, with Mandelstam invariants defined as
$s=s_{12}$ and $t=s_{13}$ in the particular case of four-point kinematics.

\subsection{Four-point and five-point tree amplitude}

\begin{figure}[h]
\centering
\includegraphics[scale=1]{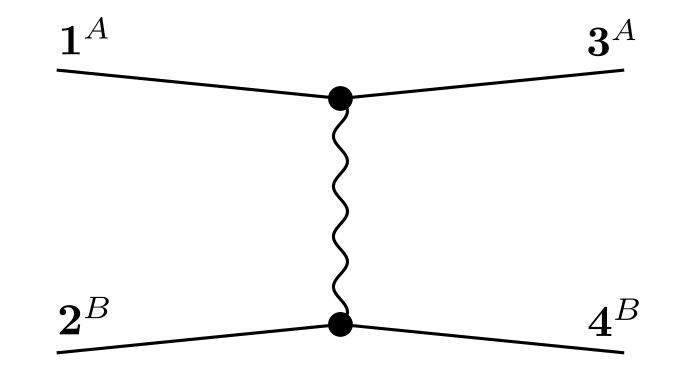}
\caption{The only Feynman diagram contributing to $\mathcal{A}_4^{(0)}({\bf 1}^A,{\bf 2}^B,{\bf 3}^A,{\bf 4}^B)$.}
\label{fig:4ptdiagram}
\end{figure}

We have only one diagram in the 4-pt case, given in Fig.~\ref{fig:4ptdiagram}. The Feynman rules give the well-known result\footnote{See for example eq.~(3.1) of \cite{KoemansCollado:2019lnh}, with $D=4$ and $\kappa =\sqrt{2} \kappa_4$.}
\begin{align}
\mathcal{A}^{(0)}_4({\bf 1}^A,{\bf 2}^B,{\bf 3}^A,{\bf 4}^B)
&= -\frac{i \kappa ^2}{2 t} \left(\frac{1}{2} t \left(-m_A^2-m_B^2+s\right)+\frac{1}{2} \left(-m_A^2-m_B^2+s\right)^2-m_A^2 m_B^2\right) .
\label{eq:4pt_twoflavors}
\end{align}

\begin{figure}[h]
\centering
\includegraphics[scale=1.75]{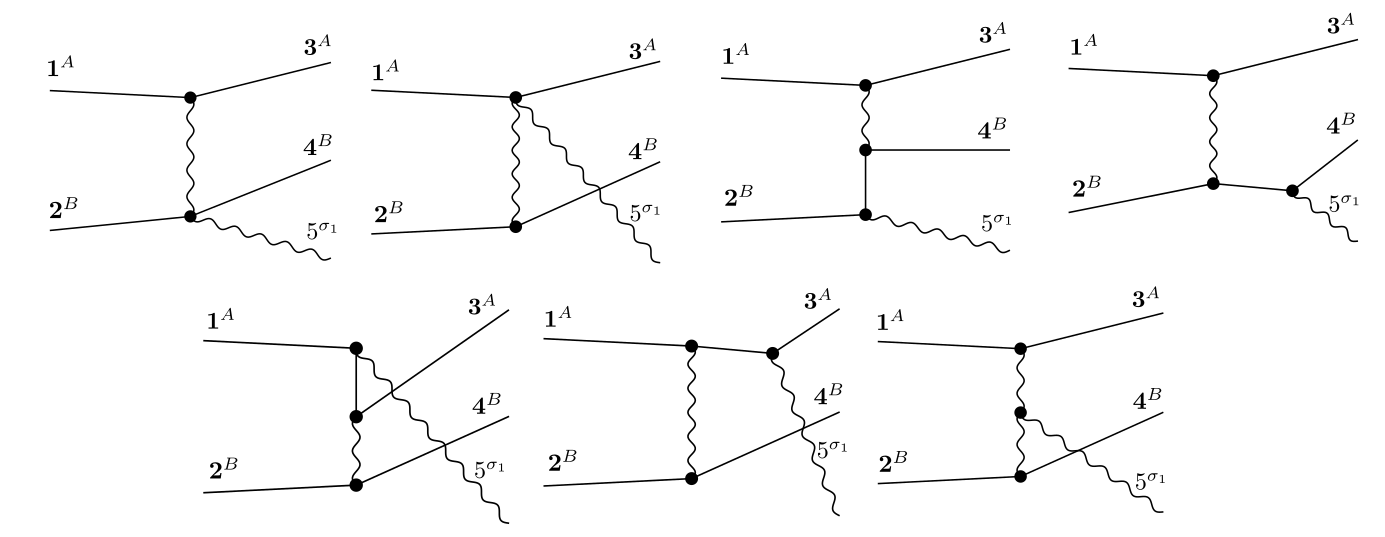}
\caption{The Feynman diagrams contributing to $\mathcal{A}_5^{(0)}({\bf 1}^A,{\bf 2}^B,{\bf 3}^A,{\bf 4}^B,5^{\sigma_1})$.}
\label{fig:5ptdiagrams}
\end{figure}

For the 5-pt amplitude, we have explicitly computed the 7 diagrams pictured in Fig.~\ref{fig:5ptdiagrams}. Notice that the first 6 diagrams are in one-to-one correspondence with the analogous calculation in scalar QED \cite{Luna2018}, while the last one is related to the graviton self-interaction.

\subsection{Six-point tree amplitude}

\begin{figure}[h]
\centering
\includegraphics[scale=1.75]{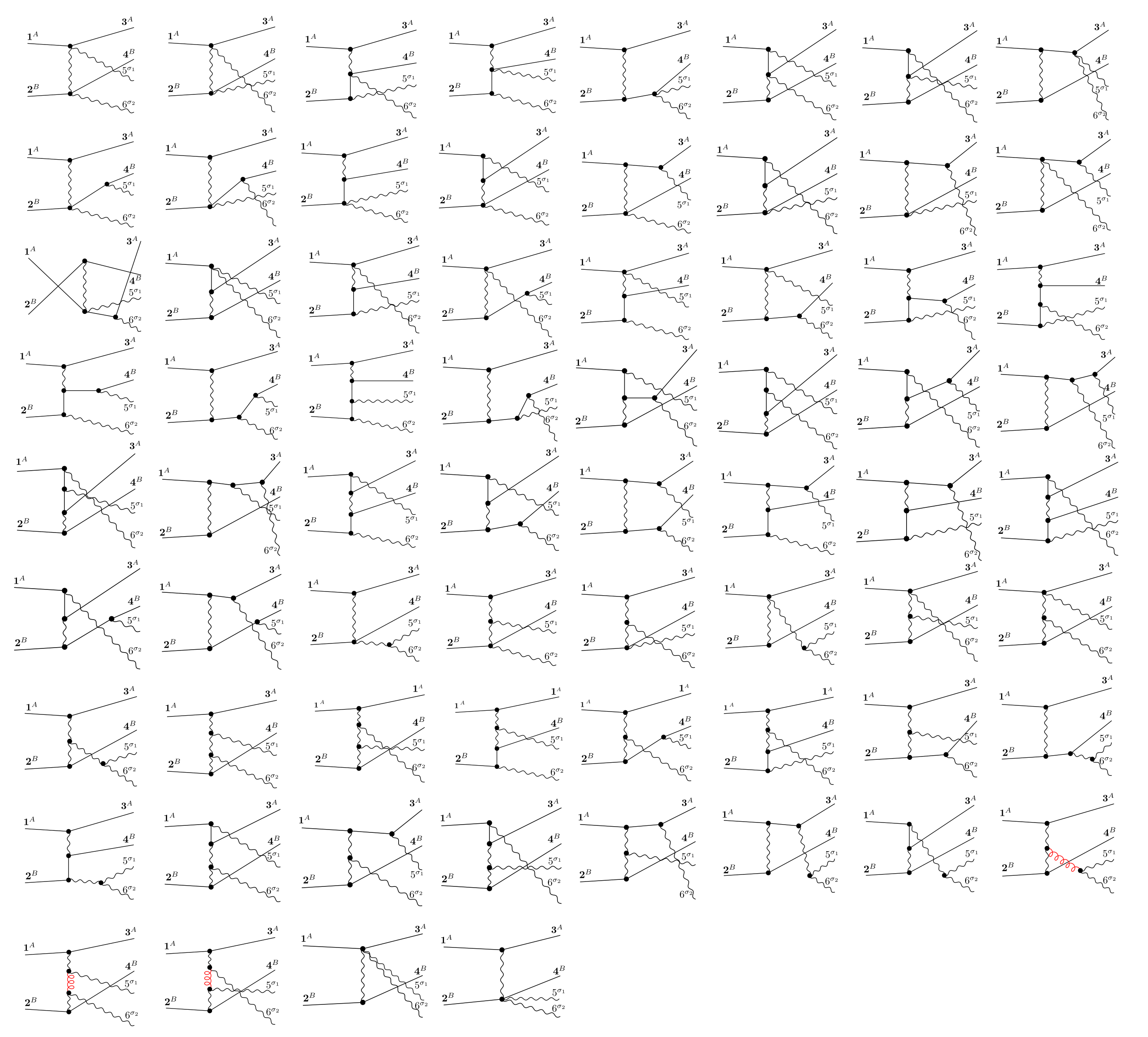}
\caption{The Feynman diagrams contributing to $\mathcal{A}^{(0)}_6({\bf 1}^A,{\bf 2}^B,{\bf 3}^A,{\bf 4}^B,5^{\sigma_1},6^{\sigma_2})$. We have highlighted in red the contribution of the auxiliary field, which is crucial to obtain the correct result.}
\label{fig:6ptdiagrams}
\end{figure}

We have computed the 68 diagrams in Fig.~\ref{fig:6ptdiagrams} for the 6-point tree amplitude. In order from the top left of the picture in Fig.~\ref{fig:6ptdiagrams}, the first 42 of these diagrams can be compared with the analogous calculation in scalar QED done in \cite{Cristofoli:2021jas}, which in particular involve the 3-point and the 4-point vertices with one matter line and one or two gravitons. The remaining 26 diagrams are classified into the following three types:
\begin{itemize}
\item 21 diagrams involving the graviton self-interaction;
\item 3 diagrams with the auxiliary field propagator;
\item 2 diagrams with a 5-point contact vertex with 3 gravitons and one matter line.
\end{itemize}

The calculation of these tree level amplitudes agrees exactly with an independent on-shell BCFW calculation presented in the next section. 

\section{Tree amplitudes from on-shell recursion relations}
\label{sec:onsh}

In this section, we compute the necessary tree-level amplitudes for the theory defined in  equation~\bref{eqn:lagrangian_GRmatter} by using an on-shell diagrammar\footnote{This terminology is taken from~\cite{tHooft:1973wag,Dunbar:2017dgp}.} to recursively construct all the amplitudes in the theory. 
A diagrammar requires basic amplitudes to serve as the atoms of the computation, and the on-shell recursive framework of BCFW~\cite{Britto2005a}. In massless theories there are straightforward arguments to construct three-point amplitudes from little-group scaling~\cite{Benincasa:2007xk,Arkani-Hamed:2017jhn}. The simplicity comes from the on-shellness of the momenta, which is maintained throughout the computation and simplifies the expressions needed as input.

We begin with a brief review of BCFW recursion, in preparation for the new shift that we will introduce to compute the 5-point tree amplitude $\mathcal{A}^{(0)}_5({\bf 1}^A,{\bf 2}^B,{\bf 3}^A,{\bf 4}^B,5^{\sigma_1})$ and to set the stage for its application to the 6-point tree amplitude  $\mathcal{A}^{(0)}_6({\bf 1}^A,{\bf 2}^B,{\bf 3}^A,{\bf 4}^B,5^{\sigma_1},6^{\sigma_2})$.

\subsection{Review of BCFW}
\label{sec:bcfw}

The basic mechanism of BCFW recursion is understood through elementary complex analysis. 
The derivation begins by introducing a complex variable $z$ and considering a linear shift in (a subset of) the momenta $p_i$ in the (yet-to-be-determined) $n$-point tree-level amplitude:
\begin{align}
	 \mathcal{A}_n^{(0)} \left( \lbrace p_i\rbrace \right) \rightarrow \mathcal{A}_n^{(0)} ( \lbrace \hat{p}_i\rbrace)\; ,
\end{align} 
where the shifted momenta are defined as
\begin{align}
	\hat{p_i} = p_i+z r_i.
\end{align}
The choice of $r_i$ corresponds to a \textit{choice of shift}.

As tree amplitudes are rational functions, we can consider $\mathcal{A}_n^{(0)}  ( \lbrace \hat{p}_i\rbrace )$ as a meromorphic function of $z$ which we denote as $\mathcal{A}_n^{(0)} (z)$. We then evaluate the contour integral
\begin{align}
	\oint_{\gamma_\infty} dz {\mathcal{A}^{(0)}_n(z)\over z} = 
	\mathcal{A}^{(0)}_n(0)+ \sum_I \underset{z=z_I}{\rm Res}\left[ {\mathcal{A}^{(0)}_n(z)\over z}\right] \; . 
	\label{eq:BCFW}
\end{align}
where the $z_I$ are the poles in the complex plane, and the integration contour $\gamma_\infty:= \lim_{R \to \infty} \gamma_R$, where $\gamma_R$ is a circular contour around the origin with radius $R$.

The choice of the vectors $r_i$ will to some extent determine the large-$z$ behavior, 
but importantly must also satisfy~\cite{Elvang2013}: 
\begin{itemize}
	\item For all $i,j$, we have $r_i\cdot r_j=0,$ which ensures linearity of deformed inverse propagators in $z$;
	\item On-shellness of the shifted momenta: $\hat{p}_i^2=-m_i^2$, which implies $r_i\cdot p_i=0$;
	\item Conservation of momentum is maintained on the shift, i.e.\ $\sum_i r_i =0$ .
\end{itemize}

With an appropriate choice of shift, and for generic kinematics, 
and the non-trivial residues on the right-hand side are thus encoded by the kinematic 
poles of the amplitude. In particular, the first condition implies that the 
poles in ${\mathcal{A}^{(0)}_n(z)}$ are simple poles. 
The residues are 
defined by the product of lower-point on-shell amplitudes in the same theory and
the scalar propagator,
\begin{align}
	\sum_I \underset{z=z_I}{\rm Res}\left[ {\mathcal{A}_n^{(0)}(z)\over z}\right] &=  
	-\sum_I \sum_{\sigma=\pm} \mathcal{A}^{(0)}_L\bigl(\lbrace\hat p_L\rbrace ,
	 \hat P_I^{\sigma} \bigr)\;{-i\over  P_I^2+m_I^2}\;
		\mathcal{A}^{(0)}_R\bigl(-\hat P_I^{-\sigma},\lbrace\hat p_R\rbrace \bigr)  \; ,
		\label{eq:poles}
\end{align}
where $L$ and $R$ stand for the ``left'' and  ``right'' amplitude in the factorization, and
\begin{align}
		P_I &= \sum_R p_R=-\sum_L p_L .
\end{align}
The momentum channels which contribute a residue are those which contain at least one shifted external momentum in both  $\lbrace\hat p_L\rbrace$ and $\lbrace\hat p_R\rbrace$, and the poles corresponding to each channel are the solutions of the linear equations
\begin{align}
	\hat P_I^2 = P_I^2 +z\sum_{i\in R} 2r_i\cdot P_I \; .
\end{align}
Note also that each pole contributes only a single residue, so partitioning into $\lbrace p_L\rbrace$ and $\lbrace p_R\rbrace$ should take into account global momentum conservation to avoid overcounting.

A ``good'' shift on $\mathcal{A}^{(0)}$ is defined as any shift for which the left-hand side of \bref{eq:BCFW} vanishes, behavior which corresponds to the vanishing of the residue at infinity, also known as the ``boundary term'',
\begin{align}
	\oint_{\gamma_\infty} dz {\mathcal{A}_n^{(0)}(z)\over z} = \underset{z\rightarrow \infty}{\rm lim}\left[ \mathcal{A}_n^{(0)} (z)\right] =0 \; .
\end{align}
For amplitudes in massless theories, it is understood what constitutes a good shift for various helicity configurations in various theories~\cite{Benincasa:2007xk,Benincasa:2007qj,ArkaniHamed:2008yf,Cohen:2010mi,Cheung2015a}.
Then, by combining \bref{eq:poles} with \bref{eq:BCFW}, we get the recursive formula
\begin{align}
	\mathcal{A}^{(0)}_n(0)= \mathcal{A}^{(0)}_n\bigl(\lbrace p\rbrace\bigr)=  
	\sum_I \sum_{\sigma=\pm} \mathcal{A}^{(0)}_L\bigl(\lbrace\hat p_L\rbrace ,
	 \hat P_I^{\sigma} \bigr)\;{-i\over  P_I^2}\;
		\mathcal{A}^{(0)}_R\bigl(-\hat P_I^{-\sigma},\lbrace\hat p_R\rbrace \bigr) .
		\label{eq:recursion}
\end{align}

Later in this section we introduce a new kind of shift which is applicable to massive legs as well. 
In particular, it will be only the first item, the on-shellness of the momenta, that needs modification to accommodate this case.

In the following section we apply BCF shifts~\cite{Britto2005b} exclusively to massless legs: they are labelled as $[i,j\ra $ 
and they modify the external legs as follows:
\begin{align}
	\hat p_i^{a\dot b} &= |\hat i]^a\la i|^{\dot b} =	\bigl(|i]+z|j]\bigr)^a\la i|^{\dot b},
	\notag \\
	\hat p_j^{a\dot b} &= |j]^a\la \hat j|^{\dot b} =	|j]^a\bigl(\la j|-z\la i|\bigr)^{\dot b}, 
\end{align}
which implies that
\begin{align}
	r_i^{a\dot b} &= -r_j^{a\dot b}= |j]^a\la i|^{\dot b} . 
\end{align}
We now proceed to apply these shifts to graviton-scalar amplitudes.

We use the spinor-helicity formalism throughout,
adopting the shorthand of~\cite{Elvang2013} whereby Feynman-slashed four-momentum is
replaced by the momentum labels with products denoted by simply concatenating momentum labels
\begin{align}
	\dots ij|X]\equiv \dots\slashed p_{i}\slashed p_j|X]   \; .
\end{align}
Differences
of momenta are similarly denoted, whilst sums of momenta are combined into an upper-case $P$:
\begin{align}
	(i-j)|X]\equiv (\slashed p_i-\slashed p_j)|X]\quad ,\quad P_{ij}= p_i+p_j\; .
\end{align}

\subsection{Building blocks of the amplitude diagrammar}
\label{sec:4ptsonsh}

\begin{figure}[ht]
	\centering{
	\includegraphics[width=0.9\textwidth]{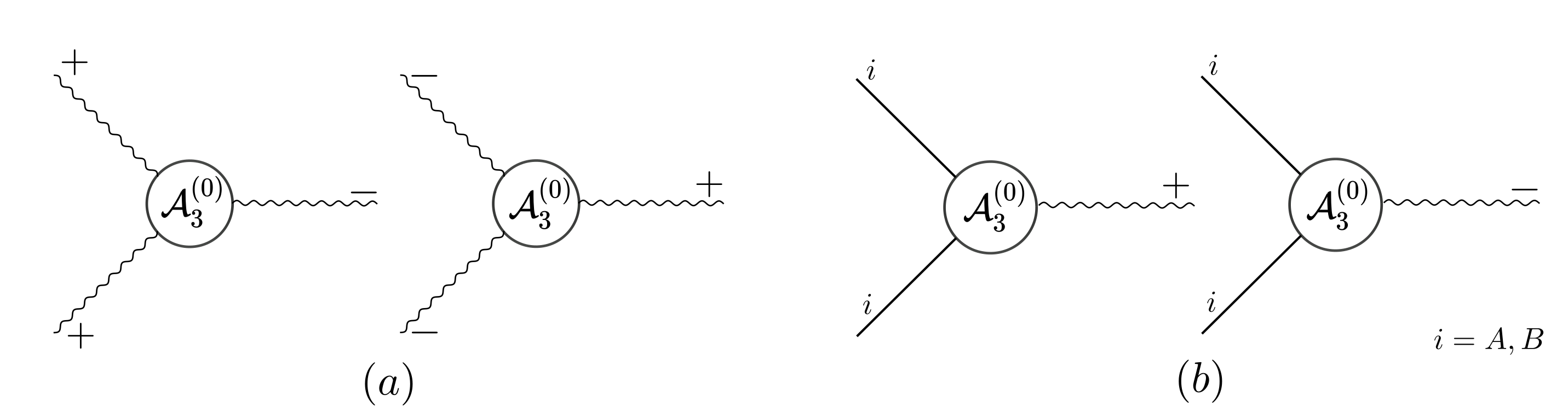}
}
	\caption[Three points]{
		Generic on-shell diagrams that are the atoms of the diagrammar, corresponding to the 
		amplitudes defined in \bref{eq:3pgrav} (resp.\ \bref{eq:3pmsiv}) for (a) (resp.\ (b)).
	}	
    \label{fig:3points}
\end{figure}

We begin by looking at amplitudes with a single flavor of massive scalar, which we pick as flavor $A$ without loss of generality. To construct these amplitudes we require pure gravity amplitudes as well as minimally  coupled graviton-scalar amplitudes. The diagrams for the three-point amplitudes needed are depicted in figure~\ref{fig:3points}.

The massless three-point graviton amplitudes are
\begin{align}
	\mathcal{A}^{(0)}_3\bigl( 1^+,2^+,3^-\bigr) &=   i\kappa{[12]^6\over [23]^2[31]^2 }, &\quad & \quad  &
	\mathcal{A}^{(0)}_3\bigl( 1^-,2^-,3^+\bigr)& =   i\kappa{\la 12 \ra^6 \over \la 23\ra^2\la 31\ra^2} ,
	\label{eq:3pgrav}
\end{align}
and the massive-scalar amplitudes are~\cite{Badger2005,Arkani-Hamed:2017jhn}
\begin{align}
	\mathcal{A}^{(0)}_3\bigl( {\bf 1}^A,{\bf 2}^A,3^+\bigr) 
	&=   i\kappa{[3|2|\chi\ra^2 \over \la 3\chi \ra^2}, &\quad & \quad &
	\mathcal{A}^{(0)}_3\bigl( {\bf 1}^A,{\bf 2}^A,3^-\bigr)
	& =   i\kappa{\la 3 | 2|\chi]^2\over [3\chi]^2} .
	\label{eq:3pmsiv}
\end{align}
where we have introduced a reference spinor $\chi$. Although it may appear as though the amplitudes~\bref{eq:3pmsiv}  depend on the choice of $\chi$, this is not the case, as long as the denominators do not vanish.

Using the amplitudes~(\ref{eq:3pgrav})--(\ref{eq:3pmsiv}) we can apply BCFW recursion to construct four-point amplitudes. Up to helicity conjugation and permutation (crossing) invariance, there are two independent configurations:
\begin{align}
	\mathcal{A}^{(0)}_4 ({\bf 1}^A,{\bf 2}^A,3^+,4^+), 	\quad  \quad  
	\mathcal{A}^{(0)}_4 ({\bf 1}^A,{\bf 2}^A,3^-,4^+) .
\end{align}
We can apply a $[3,4\ra$ shift to construct both,\footnote{For simplicity, we will suppress the $a,\dot b$ indices from here on.}
\begin{align}
	\hat p_3^{a\dot b} &= |3\ra^a[\hat 3|^{\dot b}= |3\ra^a ([3| +z[4|)^{\dot b} \; ,
	\\
	\hat p_4^{a\dot b} &= |\hat 4\ra^a[4|^{\dot b}= (|4\ra-z|3\ra )^a [4|^{\dot b} .
\end{align}
In massless theories the validity of such a shift follows directly from the scaling of the two-point propagator and polarization tensors~\cite{ArkaniHamed:2008yf,Cheung2015a}, and this analysis appears to hold for the massive case too, as it results in the correct amplitudes (see for example~\cite{BjerrumBohr2014,BjerrumBohr2019}).

\begin{figure}[t]
	\centering
		\includegraphics[width=0.9\textwidth]{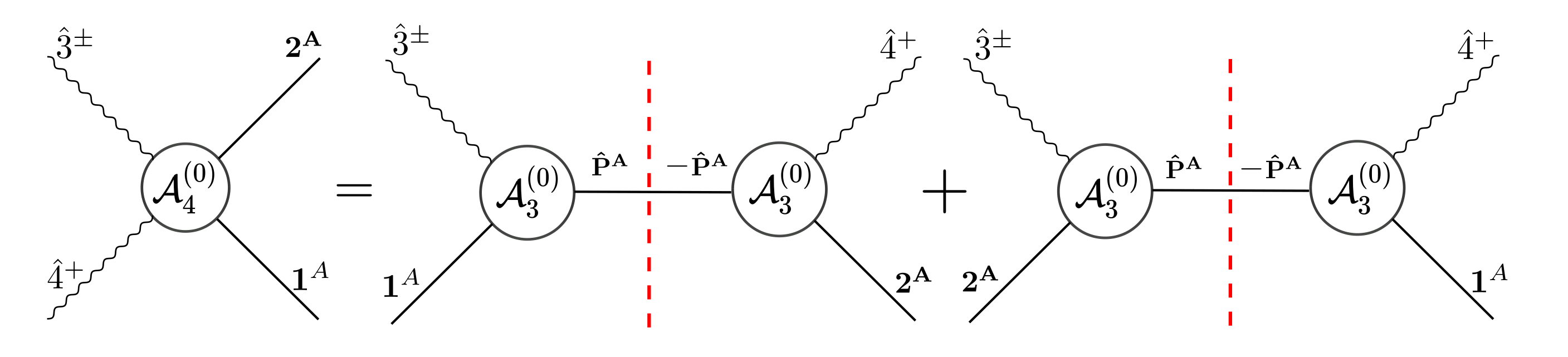}
	\caption[Four points]{The sum of factorizations of $\mathcal{A}^{(0)}_4({\bf 1}^A,{\bf 2}^A,3^{\pm},4^+\bigr)$ which is needed to reproduce the amplitude via BCFW on-shell techniques. }
    \label{fig:2Sgg}
\end{figure}

The momentum shift involves a subset of the physical poles of the theory,
\begin{align}
	\hat s_{31}-m_A^2 = 0&\Rightarrow  z=z_{31} \equiv -{[3|1|3\ra \over  [4|1|3\ra}\; ,
	\\
	\hat s_{41}-m_A^2 = 0&\Rightarrow  z=z_{41} \equiv -{[4|1|4\ra \over  [4|1|3\ra}\; ,
\end{align}
and thus the amplitudes can be reproduced by the diagrams in figure~\ref{fig:2Sgg}. 

At four points, some simple algebra reproduces a compact form of the amplitude
from the factorizations
\begin{align}
	\mathcal{A}^{(0)}_4 ({\bf 1}^A,{\bf 2}^A,3^+,4^+) &=    
	i\kappa^2{[\hat 3|1|\chi\ra^2\over \la 3 \chi\ra^2}
	{-1\over s_{31}-m_A^2}
	{[4|2|\chi\ra^2\over \la \hat 4 \chi\ra^2}+ i\kappa^2    
	{[\hat 3|2|\chi\ra^2\over \la 3 \chi\ra^2}{-1 \over s_{32}-m_A^2}
	{[4|1|\chi\ra^2\over \la \hat 4 \chi\ra^2}
	\notag \\
&=    -i\kappa^2{m_A^4[34]^2\over (s_{31}-m_A^2)\la  4 3\ra^2}
	+ (1\leftrightarrow 2) .
\end{align}
The spurious double pole cancels upon summation with the symmetric term, and the technique also gives the correct result for the mixed-helicity configuration,\footnote{In particular, the $3\leftrightarrow 4$ symmetry is restored, and the correct $s_{34}$ factorization is reproduced.}
\begin{align}
	\mathcal{A}^{(0)}_4 ({\bf 1}^A,{\bf 2}^A,3^+,4^+) &=    
	-i\kappa^2{m_A^4[34]^3\over \la 34\ra (s_{31}-m_A^2)(s_{32}-m_A^2)	}
	\label{eq:pp4S},
	 \\
	\mathcal{A}^{(0)}_4 ({\bf 1}^A,{\bf 2}^A,3^-,4^+) &=    
	i\kappa^2{[4|1|3\ra^4\over s_{34} (s_{31}-m_A^2)(s_{32}-m_A^2)}
	\label{eq:mp4S} .
\end{align}

Finally, we consider the four-point two-flavor amplitude computed from a single Feynman diagram and given in equation~\bref{eq:4pt_twoflavors}, that is equivalent to
\begin{align}
	\mathcal{A}^{(0)}_4 ({\bf 1}^A,{\bf 2}^B,{\bf 3}^A,{\bf 4}^B)=   
	{i\kappa^2\over 2s_{13}}\left({ 2(p_1\cdot p_2) (p_1\cdot p_4)+ m_A^2m_B^2}\right) .
	\label{eq:foup}
\end{align}
There are well-established on-shell constraints on the classical contribution of this 
amplitude to eikonal scattering; it consists of a single residue in the form of a product of three-point amplitudes subject to a shift prescription which defines the residue in $s_{13}$~\cite{Cachazo:2017jef}. The full QFT amplitude requires further information to fully reproduce equation~\bref{eq:foup}. Because of the simplicity of the Feynman diagram calculation, 
we treat it as a fundamental amplitude in our diagrammar, and it joins the basic
building blocks in equations~(\ref{eq:3pgrav})-(\ref{eq:3pmsiv}).

\subsection{The equal-mass shift}
\label{sec:eqmass}

\begin{figure}[h!]
	\centering
		\includegraphics[width=0.35\textwidth]{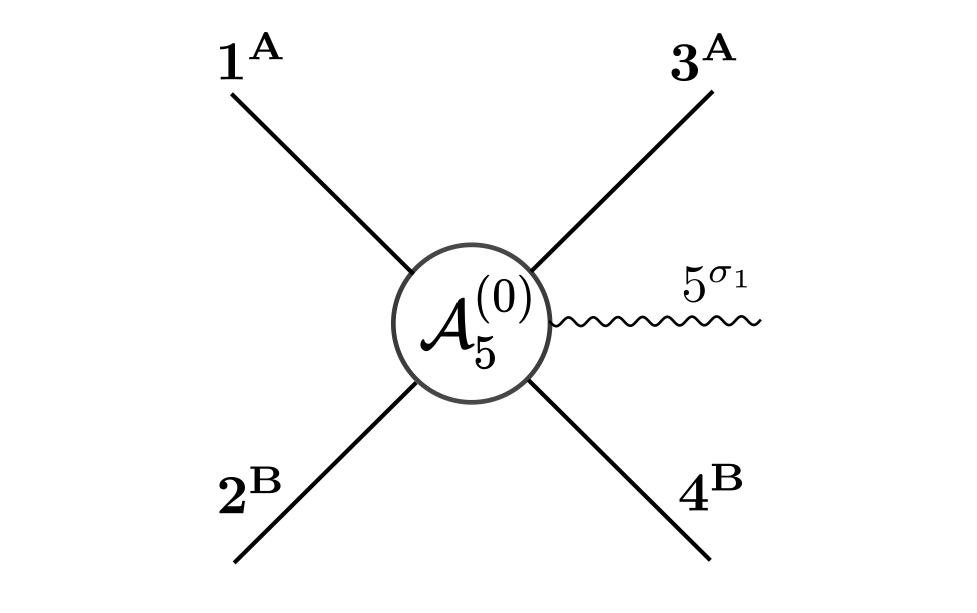}
	\caption[Five points]{The five-point tree amplitude we would like to compute with BCFW-like techniques. }
    \label{fig:5pt}
\end{figure}

The results discussed in section~\ref{sec:4ptsonsh} relied upon the presence of massless particles in the processes in question, but here we are interested in the amplitude with two massive particles with different flavors and just a single massless graviton, as depicted in figure~\ref{fig:5pt}. This raises the question of whether we can construct this amplitude with any kind of shift. In fact this is possible, but first we need to consider what actually makes on-shell recursion effective.

The principal advantage of the BCFW method is that it allows us to construct higher-point amplitudes from on-shell \textit{expressions}. When we are dealing with massless theories/particles, this also implies that the on-shell \textit{condition} for a particle is also satisfied: $\hat{p}_i^2=0$. These two statements are not completely equivalent when considering theories with equally-massive particles (particles $1$ and $3$): an on-shell expression need not be in terms of momenta and masses which satisfy the on-shell conditions
\begin{align}
	\hat p_1^2 &= \hat p_3^2= -m_A^2 
\end{align}
but can be loosened such that the mass is shifted, but by the same value for both particles:
\begin{align}
	\hat p_1^2 &= \hat p_3^2 =-\hat m_A^2 .
	\label{eq:eqshm}
\end{align}
The mass $m_A$ is thus treated like a kinematic variable rather than an invariant defining ``on-shellness''. Crucially, the equal-mass expressions now used in the recursion \textit{remain equal-mass expressions}, and the diagrammar can be used to build amplitudes in the theory just like the massless case.

This approach still requires at least one massless external particle, 
which we label particle $5$ and assume to have positive helicity, without loss of generality. 
The three-line shift that satisfies the requirements of on-shell recursion is
\begin{align}
	\hat p_5 &= |\hat 5\ra[\hat 5| = (|5\ra +z \; (1-3)|5])[5|\;, \notag \\
	\hat p_1 &= p_1 + z\;  3|5][5|\;, \notag \\
	\hat p_3 &= p_3 - z\;  1|5][5|,
\end{align}
where one can easily verify that
\begin{align}
	\hat p_1^2 = p_1^2 -z[5|13|5] 
	&= -m_A^2  +z[5|31|5] = \hat p_3^2 .
\end{align}
Thus the condition~\bref{eq:eqshm} is satisfied, and equal-mass amplitudes can be used in the recursion. Similarly to the BCFW shift, shifting the anti-holomorphic spinor $|5]$ produces a boundary term in $\mathcal A_5^{(0)}(z)$, i.e.\ it is a ``bad'' shift. From comparison with the extended-Cheung-Remmen Feynman diagram computation of section~\ref{sec:Feynman}, we confirm that the
holomorphic shift is a good shift for the five-point tree amplitude.

\subsection{Five-point tree amplitude}
\label{sec:onshfivep}

We now apply the equal-mass shift to the tree-level amplitude with two flavors of pairs of 
minimally-coupled massive particles and one graviton.

The equal-mass shift we use is 
\begin{align}
\mathcal{A}^{(0)}_5({\bf 1}^A,{\bf 2}^B,{\bf 3}^A,{\bf 4}^B, 5^+) &\rightarrow  \mathcal{A}^{(0)}_5(\hat {\bf 1}^A,{\bf 2}^B,\hat {\bf 3}^A,{\bf 4}^B, \hat 5^+) \;, \notag \\
	|\hat{5}\ra[ \hat{5}| &= \left( |5\ra +z(1-3)|5]\right) [5|\;, \notag \\
\hat p_1 &= p_1 +z\; 3|5][5|\;, \notag \\
\hat p_3 &= p_3 -z\; 1|5][5|\;, \notag \\
\hat p_1-\hat p_3 &= p_1-p_3 +z\; P_{13}|5][5| .
\label{eq:fpshifts}
\end{align}

\begin{figure}[ht]
\centering
\includegraphics[width=1\textwidth]{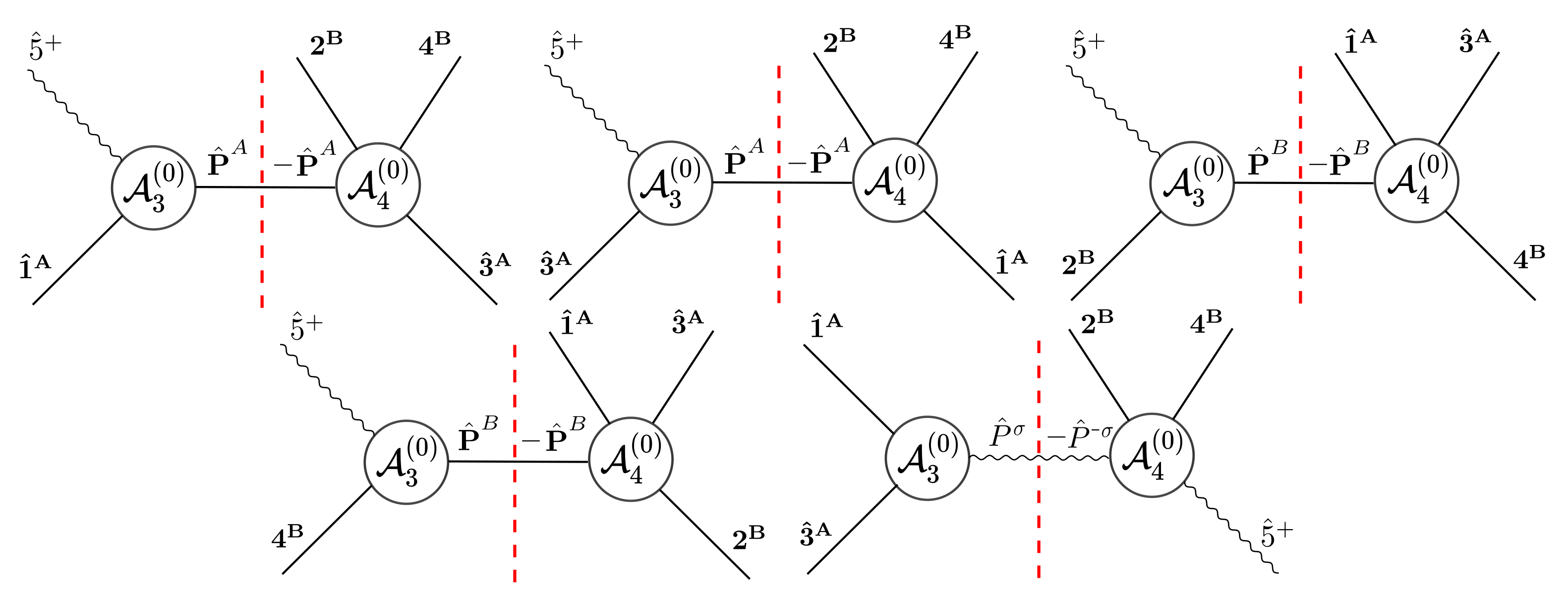}
\caption[five point factorizations]{Factorizations of the five-point tree amplitude $\mathcal{A}_5^{(0)}({\bf 1}^A,{\bf 2}^B,{\bf 3}^A,{\bf 4}^B,5^{\sigma_1})$ on the equal mass shift defined in the expressions~\bref{eq:fpshifts}.}
\label{fig:fpfacs}
\end{figure}

The factorization on the equal-mass poles are depicted in figure~\ref{fig:fpfacs}, and the shift  yields a total of five terms,
\begin{align}
\mathcal{A}^{(0)}_5 |_{51} &= 
	\mathcal{A}^{(0)}_3\left(\hat {\bf 1}^A, \hat {\bf P}^A,\hat 5^+\right)
	{i\over s_{51}-m_A^2}
	\mathcal{A}^{(0)}_4(-\hat{ \bf P}^A, {\bf 2}^B, \hat {\bf 3}^A, {\bf 4}^B) ,
	\notag \\
	\mathcal{A}^{(0)}_5 |_{53} &= \mathcal{A}^{(0)}_3\left(\hat{\bf 3}^A, \hat {\bf P}^A,\hat 5^+\right) 
	{i\over s_{53}-m_A^2} \mathcal{A}^{(0)}_4(- \hat{ \bf P}^A, {\bf 2}^B, \hat {\bf 1}^A, {\bf 4}^B) ,
	\notag \\
	\mathcal{A}^{(0)}_5|_{52} &= \mathcal{A}^{(0)}_3\left({\bf 2}^B, \hat {\bf P}^B,\hat 5^+\right) 
	{i\over s_{52}-m_B^2} \mathcal{A}^{(0)}_4(- \hat{ \bf P}^B, \hat {\bf 3}^A, {\bf 4}^B, \hat{\bf 1}^A) ,
	\notag \\
	\mathcal{A}^{(0)}_5|_{54} &= \mathcal{A}^{(0)}_3 \left({\bf 4}^B, \hat {\bf P}^B,\hat 5^+\right) 
	{i\over s_{54}-m_B^2} \mathcal{A}^{(0)}_4 (-\hat{ \bf P}^B, \hat {\bf 3}^A, {\bf 2}^B, \hat{\bf 1}^A), 
	\notag \\
	\mathcal{A}^{(0)}_5|_{13} &= \sum_{\sigma=\pm} \mathcal{A}^{(0)}_3 \left(\hat {\bf 1}^A,\hat {\bf 3}^A, \hat P^{\sigma}\right) 
	{i\over s_{13}} \mathcal{A}^{(0)}_4( {\bf 2}^B, {\bf 4}^B, \hat 5^+,-\hat P^{-\sigma}),
\label{eq:fac5pt}
\end{align}
where the factorizations correspond to residues at the following poles:
\begin{align}
&z_{51}={[ 5|1|5\ra \over [5|13|5]} \,,\;  \quad  &z_{53}=& {[ 5|3|5\ra \over [5|13|5]} \,,\nonumber \\
&z_{52}= -{[5|2|5\ra \over [5|2(1-3)|5]} \,,\;  \quad &z_{54}=& -{[5|4|5\ra \over [5|4(1-3)|5]} \,, \nonumber \\
&z_{13} = {s_{13}\over [5|P_{24}(1-3)|5]} \; .
\end{align}
It is convenient to organize the calculation in terms of the variables\footnote{Note that because the momenta are all incoming, the $K_i$ are {\it not} momentum transfers.}
\begin{align}
K_A := p_1-p_3, \qquad K_B := p_2-p_4,
\end{align}
which are antisymmetric under the exchange of the corresponding pair of momenta. 
Each residue in  \eqref{eq:fac5pt} yields an expression containing spurious poles, which are not present in the full amplitude. For example the $P_{52}$ factorization gives
\begin{align}
	\mathcal{A}^{(0)}_5\bigr|_{52}= &{i \kappa^3} {[5|2K_A|5]^3[5|13|5]^2\over x_{51|53}^2}{-1\over s_{52}-m_B^2} \times 
	\notag \\
	&{ 2(p_1\cdot p_4+z_{52}[5|43|5]) ( p_3\cdot p_4-z_{52}[5|41|5])+ (m_A^2+z_{52}[5|13|5])m_B^2\over 2x_{52|13}} ,
	 \label{eq:m52a}
\end{align}
with the spurious poles $x_{ij|kl}$ proportional to denominator factors evaluated at other 
residues
\begin{align}
	x_{ij|kl}= [5|P_{ij}K_A|5][5|P_{kl}K_A|5](z_{ij}-z_{kl}) .
\end{align}

Through algebraic manipulations the spurious poles in the full expression can be cleared, and 
the amplitude can be symmetrized in $K_A$ and $K_B$. The final expression is
\begin{align}
	\mathcal{A}^{(0)}_5&\left({\bf 1}^A,{\bf 2}^B,{\bf 3}^A,{\bf 4}^B,5^+\right)=
	{i\kappa^3 \over 8} \biggl(\biggl[{-p_4\cdot p_2[5|13|5]^2\over s_{24}(s_{51}-m_A^2)(s_{53}-m_A^2)}+
	{[5|K_AK_B|5]^2-8[5|13|5]^2\over 16 s_{13}s_{24}}
	&
	\notag \\
	&+{(m_A^2+m_B^2)[5|13|5]\bigl(2(s_{13}-s_{24})[5|13|5]+ [5|K_B|5\ra[5|K_AK_B|5]\bigr)\over 8 
	(s_{51}-m_A^2)(s_{53}-m_A^2)(s_{52}-m_B^2)(s_{54}-m_B^2)}
	\notag \\
	&-{K_A\cdot K_B(s_{24}-s_{13})^2[5|13|5]\bigl(4s_{24}[5|42|5]-[5|K_A|5\ra [5|K_AK_B|5]\bigr) \over 
	32s_{13}s_{24}(s_{51}-m_A^2)(s_{53}-m_A^2)(s_{52}-m_B^2)(s_{54}-m_B^2)}
	\notag \\
	&-{K_A\cdot K_B[5|42|5]\bigl([5|K_B|5\ra[5|K_AK_B|5]-4(s_{13}+s_{24})[5|13|5]\bigr)\over 
	8s_{13}s_{24}(s_{52}-m_B^2)(s_{54}-m_B^2)}
	\notag \\ 
	&-{K_A\cdot K_B[5|K_A|5\ra[5|K_B|5\ra \bigl([5|K_AK_B|5]^2-8[5|42|5]^2\bigr)\over 
	64s_{13}(s_{51}-m_A^2)(s_{53}-m_A^2)(s_{52}-m_B^2)(s_{54}-m_B^2)}
	\notag \\
	&+(\tr(\slashed K_A\slashed K_B\slashed K_A\slashed K_B) +2[5|K_A|5\ra^2+2[5|K_B|5\ra^2-2s_{13}^2-2s_{24}^2)
	\notag \\
	&\times \biggl({[5|K_A|5\ra[5|K_B|5\ra [5|13|5][5|42|5]\over 
	64 s_{13}s_{24}(s_{51}-m_A^2)(s_{53}-m_A^2)(s_{52}-m_B^2)(s_{54}-m_B^2)}
	\notag \\
	&\qquad +{[5|42|5](2(s_{13}-s_{24})[5|42|5]-[5|K_A|5\ra[5|K_AK_B|5])\over 
	64 s_{13}(s_{51}-m_A^2)(s_{53}-m_A^2)(s_{52}-m_B^2)(s_{54}-m_B^2)}
	\biggr)\biggr]+ 
\left[\bigl(1,3,K_A) \leftrightarrow \bigl(2,4,K_B\bigr)\right]\biggr) ,
	\label{eq:M5final}
\end{align}
and the negative helicity case is obtained simply by switching the square brackets for angle brackets.  We find perfect agreement with the Feynman diagram calculation from section~\ref{sec:Feynman}  when tested on rational kinematic points.

\subsection{Six-point tree amplitude}

\begin{figure}[h]
\centering
\includegraphics[width=1\textwidth]{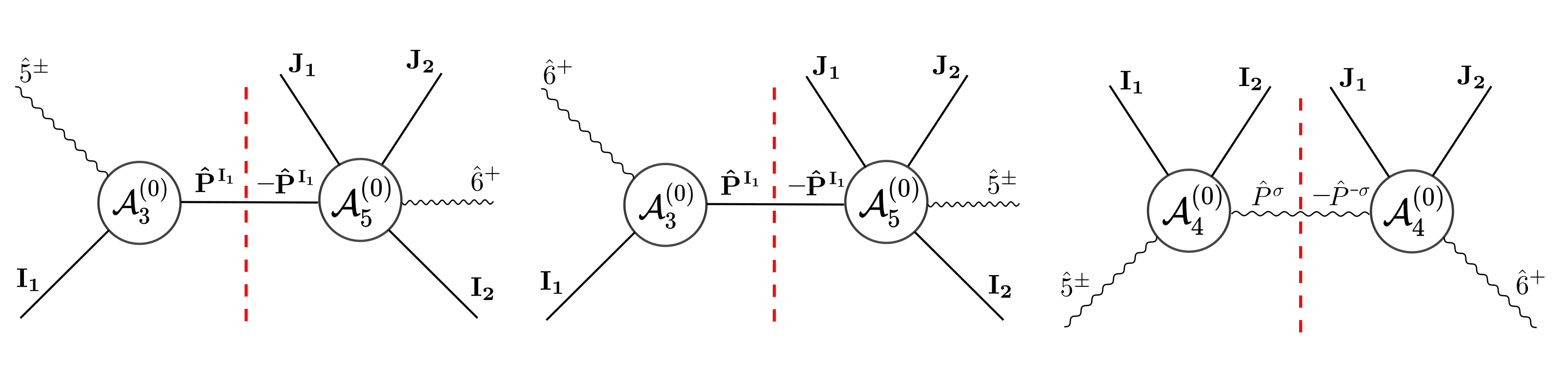}
\caption[Six point]{Schematic representation (up to crossing symmetry) of the three independent factorizations which are relevant for the construction of the six-point tree amplitude $\mathcal{A}_6^{(0)}({\bf 1}^A,{\bf 2}^B,{\bf 3}^A,{\bf 4}^B,5^{\sigma_1},6^{\sigma_2})$ from the BCFW shift.}
\label{fig:4Sggbcf}
\end{figure}

The six-point tree amplitude can be computed using a standard BCFW shift $[5,6\ra $,\footnote{The three-line shift used in section~\ref{sec:eqmass} produces more factorisations making the form less efficient. Moreover the presence of a boundary term in the same-helicity case restricts its application to generic configurations.} where we consider
\begin{align}
	\mathcal A_6^{(0)}( {\bf 1}^A, {\bf 2}^B, {\bf 3}^A,{\bf 4}^B ,\hat 5^{\pm},\hat 6^+)\;  ,
\end{align}
which generates 10 factorization diagrams. All of these are of the general types of factorizations are shown in figure~\ref{fig:4Sggbcf}. We make use of the permutation invariance of the scalar particle by defining $\lbrace \bf{I}_1,{\bf I}_2 \rbrace = P\left( \lbrace {\bf 1}^A, {\bf 3}^A\rbrace \right) $ or $\lbrace \bf{I}_1,{\bf I}_2 \rbrace =  P\left( \lbrace {\bf 2}^B, {\bf 4}^B\rbrace \right) $, with the complement set labelled as ${\bf J}_i$. 
There are four factorizations for each of the left and middle diagrams and two for the last, 
giving a total of ten residue contributions to the amplitude.

\begin{align}
\mathcal{A}^{(0)}_6|_{5I_1} &= \mathcal{A}^{(0)}_3\left({\bf I_1}, \hat{\bf P}^{\bf I_1},\hat 5^{\pm}\right) {i\over s_{5I_1}-m_I^2} \mathcal{A}^{(0)}_5(-\hat{ \bf P}^{\bf I_1}, {\bf J}_1, {\bf I}_2, {\bf J}_2, \hat 6^+)
\label{eq:type5I}
 \\
\mathcal{A}^{(0)}_6|_{6I_1} &= \mathcal{A}^{(0)}_3\left({\bf I_1}, \hat {\bf P}^{\bf I_1},\hat  6^+\right) {i\over s_{6I_1}-m_I^2} \mathcal{A}^{(0)}_5(-\hat{ \bf P}^{\bf I_1}, {\bf J}_1,  {\bf I}_2, {\bf J}_2,\hat 5^\pm) 
\label{eq:type6I}
\\
\mathcal{A}^{(0)}_6|_{I_1I_25} &= \sum_{\sigma=\pm} \mathcal{A}^{(0)}_4\left({\bf I_1},{\bf I_2}, \hat 5^{\pm}, \hat P^{\sigma}\right) {i\over s_{I_1I_25}} \mathcal{A}^{(0)}_4\left({\bf J_1},{\bf J_2}, \hat 6^{+}, -\hat P^{-\sigma}\right) 
\label{eq:typeII5}
\end{align}

We confirm numerically the vanishing of the boundary (large-$z$) terms from the Feynman-diagram 
expression.\footnote{This occurs diagram by diagram for the mixed helicity case, but there are non-trivial cancellations in the case of the all-plus-helicity configuration.}  Moreover, we have verified that the reproduction of the amplitude, as the Feynman-diagram and on-shell calculations produce the same result on all (rational) numerical points tested.

\section{The classical limit of scattering amplitudes with radiation: graviton interference is a quantum effect}
\label{sec:classical_limit}

In this section, we use the explicit calculation of the six-point tree amplitude $\mathcal{A}_6^{(0)}$  of the previous section to prove the coherence of the emitted semiclassical radiation field up to order $\mathcal{O}(G^4)$ for radiative observables. Moreover, assuming coherence to all orders as suggested by the arguments of section~\ref{sec:gravstats}, we derive an infinite set of non-trivial relations between unitarity cuts in the classical limit. Those are relevant for the calculation of physical radiative observables, such as the waveform or the total linear and angular momentum emitted by the gravitons, because they suggest that only the 5-pt amplitude is required for the classical calculation and all the higher multiplicity amplitudes are not explicitly needed.

In order to take the classical limit, we follow the rules established in \cite{Kosower:2018adc}. We express the massless momenta in terms of their wavenumbers and the momentum transfers of (\ref{eq:momentumtransfers}),
\begin{align}
k_i = \hbar \bar{k}_i \quad \textrm{for~}i=1,2,3,\ldots \, ; \qquad q_j = \hbar \bar{q}_j, \quad  w_j = \hbar \bar{w}_j \qquad  \, \textrm{for~}j=1,2 ;
\label{eqn:momenta_resc1}
\end{align}
and we use the parametrization of the massive momenta from (\ref{eq:symm-mom}), which define the classical trajectory. They are therefore associated to classical velocities $v_A$ and $v_B$,
\begin{align}
p_j = \tilde{m}_j v_j, \qquad \tilde{m}_j^2 = m_j^2 - \hbar^2 \frac{\bar{q}^2}{4} \qquad  \, \textrm{for~}j=A,B \, .
\label{eqn:momenta_resc2}
\end{align}
Note that in section~\ref{sec:onsh} we used  notation which was more compact for the purposes of computing the amplitudes. We can translate to the notation introduced earlier in  equation~(\ref{eqn:probability2}) by noticing that
\begin{align}
	P_{13} = -w_1 \; ,\; P_{24} = -w_2 \; .
\end{align}
Crucially, we also need to restore the powers of $\hbar$ in the coupling as 
\begin{align}
\kappa \to \frac{\kappa}{\sqrt{\hbar}} .
\label{eqn:coupling_resc}
\end{align}
We use these equivalences to infer the $\hbar$ scaling of the amplitudes. We begin by extracting the leading classical scaling of the five-point and six-point amplitude, and we then discuss the consequences of coherence for classical radiative observables.

\subsection{Classical limit of the five-point tree amplitude}
We begin by computing the classical limit of the five-point tree amplitude, which was given previously in \cite{Luna2018,Mogull:2020sak} by an equivalent large mass expansion. An interesting alternative derivation can be made in supergravity theory by using the Kaluza-Klein compactification of amplitudes of massless particles in five dimensions, by taking advantage of a straightforward application of double copy \cite{Bautista:2019evw}.

The manifestly gauge invariant expression for $\mathcal{A}_5^{(0)}$ given in equation~\bref{eq:M5final} can easily be written in terms of the polarization tensor for the graviton through the identification
\begin{align}
f^{\mu\nu} = p_5^\mu \varepsilon_5^\nu- p_5^\nu\varepsilon_5^\mu =[5|\gamma^\mu\gamma^\nu|5] \stackrel{\hbar \to 0}{\sim} \hbar , 
	\label{eq:fdef}
\end{align}
and the following scalings also hold:
\begin{align}
 P_{13}, P_{24} &\stackrel{\hbar \to 0}{\sim} \hbar ,\qquad \qquad K_A^\mu\; , \;  K_B^\nu \stackrel{\hbar \to 0}{\sim}\hbar^0, \qquad \qquad  s_{13}\; ,\; s_{24}\stackrel{\hbar \to 0}{\sim} \hbar^2,  \notag \\
s_{51}-m_A^2 &=-2p_5\cdot p_1 \; , \qquad s_{52}-m_B^2 =-2p_5\cdot p_2  \stackrel{\hbar \to 0}{\sim}  \hbar .
\label{eq:qscalK}
\end{align}
Moreover, we can safely neglect the quantum shift in the masses $\tilde{m}_j \stackrel{\hbar \to 0}{=} m_j$. Using equation~\bref{eq:qscalK}, we can simply apply power counting to each of the terms in expression~\bref{eq:M5final}. We deduce that, upon including the contribution from $\kappa$, the terms which contribute to leading behavior as $\hbar \to 0 $ are
\begin{align}
\mathcal{A}^{(0)}_5 &({\bf 1}^A,{\bf 2}^B,{\bf 3}^A,{\bf 4}^B,5^{+}) 
	\stackrel{\hbar \to 0}{\sim}  \notag \\
	&\frac{i \kappa^3}{64} \biggl(\biggr[{[5|K_AK_B|5]^2\over 2 s_{13}s_{24}}-{K_A\cdot K_B[5|42|5][5|K_B|5\ra[5|K_AK_B|5]\over s_{13}s_{24}(s_{52}-m_B^2)(s_{54}-m_B^2)} \notag \\ 
	&-{K_A\cdot K_B[5|K_A|5\ra[5|K_B|5\ra[5|K_AK_B|5]^2\over 
	8s_{13}(s_{51}-m_A^2)(s_{53}-m_A^2)(s_{52}-m_B^2)(s_{54}-m_B^2)} \notag \\
	&+ \tr(\slashed K_A\slashed K_B\slashed K_A\slashed K_B) \biggl({[5|K_A|5\ra[5|K_B|5\ra [5|13|5][5|42|5]\over 
	8 s_{13}s_{24}(s_{51}-m_A^2)(s_{53}-m_A^2)(s_{52}-m_B^2)(s_{54}-m_B^2)} \notag \\
	&-{[5|42|5][5|K_A|5\ra[5|K_AK_B|5]\over 8 s_{13}(s_{51}-m_A^2)(s_{53}-m_A^2)(s_{52}-m_B^2)(s_{54}-m_B^2)} \biggr)\biggr]
	+ \left[\bigl(1,3,K_A) \leftrightarrow \bigl(2,4,K_B\bigr)\right]\biggr).
\label{eq:M5cll}
\end{align}
We can make the following replacements in order to match the notation in~\cite{Luna2018} at leading order in the classical expansion,\footnote{Only for this case, we use an asymmetric parametrization of the external momenta in terms of the classical velocities just to show the agreement with the literature.} 
\begin{align}
	p_3 \stackrel{\hbar \to 0}{\sim} - m_A v_A , \qquad \qquad \qquad  \qquad\qquad \qquad & \quad p_4 \stackrel{\hbar \to 0}{\sim} - m_B v_B ,
	\notag \\
	s_{13} \stackrel{\hbar \to 0}{\sim} -q_1^2 , \qquad \qquad \qquad \qquad \qquad \quad\quad\,\,\,\,\,\,\,\,\,& \quad s_{24} \stackrel{\hbar \to 0}{\sim} -q_2^2 ,
	\notag \\ 
	[5|13|5] \stackrel{\hbar \to 0}{\sim}  m_A f_{\mu\nu}v_A^\mu q_1^\nu , \qquad\,\, \qquad \qquad   \qquad & \quad [5|42|5] \stackrel{\hbar \to 0}{\sim} -m_B f_{\mu\nu}v_B^\mu q_2^\nu ,
	\notag \\	
	\quad [5|K_A|5\ra \stackrel{\hbar \to 0}{\sim} -4m_A k\cdot v_A ,
	\quad \qquad \qquad \,\,\,\qquad  &
	\quad [5|K_B|5\ra \stackrel{\hbar \to 0}{\sim} -4m_B k\cdot v_B , \qquad \qquad
	\notag \\
	(-s_{51}+m_A^2), (s_{53}-m_A^2) \stackrel{\hbar \to 0}{\sim} 2m_A k\cdot v_A ,  \quad & 
	\quad (-s_{52}+m_B^2), (s_{54}-m_B^2) \stackrel{\hbar \to 0}{\sim} 2m_B k\cdot v_B ,
	\notag \\
	[5|K_AK_B|5] \stackrel{\hbar \to 0}{\sim} -4m_Am_Bf_{\mu\nu}v_A^\mu  v_B^\nu  ,\qquad \quad  \, &\quad K_A\cdot K_B \stackrel{\hbar \to 0}{\sim} 4m_Am_Bv_A\cdot v_B\;.
\end{align}
This implies that the leading order behaviour of the five-point tree amplitude is of order $\hbar^{-7/2}$. As we will see later, this will imply that the amplitude contributes to the total classical energy emitted in gravitational waves. In particular, we get
\begin{align}
&\mathcal{A}^{(0)}_5 \left(m_A v_A, m_B v_B,\hbar q_1 - m_A v_A,\hbar q_2 - m_B v_B,\hbar k\right)\biggr|_{\hbar^{-\frac{7}{2}}} \nonumber \\
&= 
	\frac{i \kappa^3}{4} {m_A^2m_B^2f_{\mu\nu}f_{\rho\sigma}\over q_1^2q_2^2}\biggl[v_A\cdot v_B\left({q_1^\mu v_A^\nu\over k\cdot v_A}-{q_2^\mu v_B^\nu\over k\cdot v_B} \right)v_A^\rho v_B^\sigma +
	{(q_1^2+q_2^2)v_A\cdot v_B v_A^\mu v_B^\nu v_A^\rho v_B^\sigma\over 2k\cdot v_Ak\cdot v_B}& \notag \\ 
	&+v_A^\mu v_B^\nu v_A^\rho v_B^\sigma +\tr(\slashed v_A\slashed v_B\slashed v_A\slashed v_B) \biggl(-{ v_A^\mu q_1^\nu v_B^\rho q_2^\sigma \over 4k\cdot v_Ak\cdot v_B}-
	{q_2^2v_A^\mu v_B^\nu v_B^\rho q_2^\sigma \over 8 k\cdot v_A (k\cdot v_B)^2}
	- {q_1^2v_B^\mu v_A^\nu v_A^\rho q_1^\sigma \over  8k\cdot v_B (k\cdot v_A)^2} \biggr) \biggl],
	\label{eq:M5cl2}
\end{align}
where $f_{\mu\nu}f_{\rho\sigma}$ is proportional to the linearized Riemann tensor and can be expressed
in terms of the polarization tensor $\varepsilon_{\mu\nu}$ 
\begin{align}
	f_{\mu\nu}f_{\rho\sigma} &= 2R_{\mu\nu\rho\sigma}= k_\mu k_\rho \varepsilon_{\nu\sigma}
	+k_\nu k_\sigma\varepsilon_{\mu\rho}
	-k_\mu k_\sigma\varepsilon_{\nu\rho}
	-k_\nu k_\rho\varepsilon_{\mu\sigma}
	\notag \\
	&= \varepsilon^{\alpha \beta} \left(k_\mu k_\rho\eta_{\nu\alpha}\eta_{\sigma\beta}
	+k_\nu k_\sigma\eta_{\mu\alpha}\eta_{\rho\beta}
	-k_\mu k_\sigma\eta_{\nu\alpha}\eta_{\rho\beta}
	-k_\nu k_\rho\eta_{\mu\alpha}\eta_{\sigma\beta}\right)\; .
\end{align}
Upon substituting the relation
\begin{align}
	\tr(\slashed v_A\slashed v_B\slashed v_A\slashed v_B)  = 8(v_A\cdot v_B)^2 -4 
\end{align}
the amplitude can thus be expressed 
\begin{align}
	\mathcal{A}^{(0)}_5 &\left(m_A v_A, m_B v_B,\hbar q_1 - m_A v_A,\hbar q_2 - m_B v_B,\hbar k\right)\biggr|_{\hbar^{-\frac{7}{2}}}\notag \\
	&= \frac{i \kappa^3}{4}
	{m_A^2m_B^2\varepsilon_{\mu\nu}\over q_1^2q_2^2}\biggl[
	(k\cdot v_A)^2v_B^\mu v_B^\nu+(k\cdot v_B)^2v_A^\mu v_A^\nu
	-2k\cdot v_Ak\cdot v_B v_A^\mu v_B^\nu \notag \\
	 &+v_A\cdot v_B\biggl(
	{(k\cdot v_A)^2 q_2^2v_B^\mu v_B^\nu+(k\cdot v_B)^2 q_1^2 v_A^\mu v_A^\nu
	\over 2k\cdot v_A\; k\cdot v_B}
	 + k\cdot v_A \; q_2^\mu v_B^\nu+k\cdot v_B\; q_1^\mu v_A^\nu -(q_1^2+q_2^2)v_A^\mu v_B^\nu
	\biggr) \notag \\ 
	&+\left(2(v_A\cdot v_B)^2 -1\right) \biggl(
	 { \left(q_1^2k\cdot v_Bv_A^\mu- q_2^2k\cdot v_Av_B^\mu\right) \left(q_1^\nu-q_2^\nu \right)\over 
	 2k\cdot v_Bk\cdot v_A} -{ q_1^\mu q_2^\nu }+
	 \notag \\ &\hspace{50pt}
	+{ (q_1^2-q_2^2)^2 \left((k\cdot v_A)^2q_2^2 v_B^\mu v_B^\nu +(k\cdot v_B)^2q_1^2v_A^\mu v_A^\nu\right)\over 4(k\cdot v_A)^2(k\cdot v_B)^2}
	-{(q_1^2-q_2^2)^2 v_A^\mu v_B^\nu \over 4k\cdot v_A k\cdot v_B} \biggr) \biggl] ,
	\label{eq:M5cl3}
\end{align}
which matches the result in~\cite{Luna2018} analytically.

\subsection{Classical limit of the six-point tree amplitude}

To compute the leading terms of the classical expansion of $\mathcal{A}^{(0)}_6$, we directly extract the $\hbar$ scaling of the BCFW residues in equations~(\ref{eq:type5I})--(\ref{eq:typeII5}). In the following, we will use explicitly the rules extracted in equations (\ref{eqn:momenta_resc1}),  (\ref{eqn:coupling_resc}), and (\ref{eq:qscalK}). First we consider the terms which originate from the factorizations of the general type~(\ref{eq:type5I}),
\begin{align}
\mathcal{A}^{(0)}_6|_{5I_1} &= \mathcal{A}^{(0)}_3\left({\bf I_1}, \hat{\bf P}^{\bf I_1},\hat 5^{\pm}\right) 
	{i\over s_{5I_1}-m_I^2} \mathcal{A}^{(0)}_5(-\hat{ \bf P}^{\bf I_1}, {\bf J}_1, {\bf I}_2, {\bf J}_2, \hat 6^+). 
\label{eq:bcfac}
\end{align}
For the scaling of the three-point amplitude $\mathcal{A}^{(0) }_3({\bf I_1}, \hat {\bf P}_I,\hat 5^{\pm})$, we first note that a shift in momenta does not modify the $\hbar$ scaling,
\begin{align}
	\hat p_5 = p_5 + z_{5I_1}|5\ra[6| \to \hbar \hat{\bar{p}}_5 ,
	\label{eq:p5_shift_quantum}
\end{align}
which can be seen from the fact that $z_{5I_1}$ takes the form
\begin{align}
	z_{5I_1}|5\ra[6| = {2p_5\cdot p_{I_1}\over [6|I_1|5\ra}|5\ra[6|
\end{align}
so that it scales in the same way as $p_5$.
We can thus rearrange the amplitude to extract the scaling:
\begin{align}
	\mathcal{A}_3^{(0) }\left(\hat {\bf I}_1, \hat {\bf P}^{\bf I_1},\hat 5^{+}\right) 
	&= i \kappa {[\hat 5|\hat P_I|\chi\ra^2 \over \la 5\chi\ra^2}
	\notag \\
	&=i \kappa {[\hat 5|\hat P_I\chi|\hat 5]^2 \over (2p_5\cdot p_\chi)^2} \notag \\
	&=i \kappa {1\over (2p_5\cdot p_\chi)^2}\hat f_{ 5}^{\mu\nu} \hat f_{5}^{\rho\sigma} 
	\hat P_\mu P_\rho (p_{\chi})_{\nu} (p_{\chi})_{\sigma}
	\notag \\
	&\stackrel{\hbar \to 0}{\sim}  \hbar^{-\frac{1}{2}} \; ,
\end{align}
where $\hat f_5\equiv f$ as defined in equation~(\ref{eq:fdef}), but in terms of shifted momenta and polarization vectors.
The shifted five-point amplitude inherits the $\hbar$ scaling of equation~\bref{eq:M5cl2},
\begin{align}
	\mathcal{A}^{(0) }_5(-\hat{ \bf P}^{\bf I_1}, {\bf J}_1, {\bf I}_2, {\bf J}_2, \hat 6^+) 
	\stackrel{\hbar \to 0}{\sim} \hbar^{-{7\over 2}}.
\end{align}
Thus upon including the contribution from the pole, each term of the form~\bref{eq:bcfac} has the leading scaling behavior
\begin{align}
	\mathcal{A}_6^{(0) }|_{6I_1} \stackrel{\hbar \to 0}{\sim} \hbar^{-5}.
\end{align}

We now show how taking only the leading-classical term trivializes the kinematics. Using equation~\bref{eq:p5_shift_quantum} we have
\begin{align}
	\hat P^{ I_1} = -p_{I_1}- \hat p_5= -p_{I_1} + \mathcal{O} (\hbar) \; ,
\end{align}
and we can make the statement
\begin{align}
	\mathcal{A}^{(0) }_5(-\hat{ \bf P}^{\bf I_1}, {\bf J}_1, {\bf I}_2, {\bf J}_2, \hat 6^+)\biggr|_{\hbar^{-{\frac{7}{2}}}}
	=\mathcal{A}^{(0) }_5({\bf I}_1, {\bf J}_1, {\bf I}_2, {\bf J}_2, \hat 6^+)\biggr|_{\hbar^{-{\frac{7}{2}}}} \; .
\end{align}
This is not the only simplification in the leading classical limit. We also observe that from $p_{I_1}=-p_{I_2}+\mathcal{O}(\hbar)$ we have
\begin{align}
	z_{I_15} = {[5|I_1|5\ra\over [6|I_1|5\ra} = {[5|I_2|5\ra\over [6|I_2|5\ra} +\mathcal{O}(\hbar) .
\end{align}
Thus both the $\mathcal{A}^{(0) }_3$ and $\mathcal{A}^{(0) }_5$ factors in~\bref{eq:bcfac} are invariant under the $I_1\leftrightarrow I_2$ permutation. On the other hand the pole factor
\begin{align}
{1\over s_{I_15} - m_I^2} = {1\over [5|I_1|5\ra} = -{1\over [5|I_2|5\ra}
\end{align}
has the opposite sign under the $I_1\leftrightarrow I_2$ switch, so these contributions cancel pairwise, giving 
\begin{align}
\mathcal{A}_6^{(0) }|_{5I_1}+\mathcal{A}_6^{(0) }|_{5I_2} \stackrel{\hbar \to 0}{\sim} \hbar^{-4}\; .
\end{align}
An identical argument for the terms of type~(\ref{eq:type6I}) also gives
\begin{align}
\mathcal{A}_6^{(0) }|_{6I_1}+\mathcal{A}_6^{(0) }|_{6I_2} \stackrel{\hbar \to 0}{\sim} \hbar^{-4}\; .
\end{align}
So the permutation invariance naturally leads to a drop in inverse-$\hbar$ scaling.
 
Finally, describing the scaling of terms of the type~(\ref{eq:typeII5}),
\begin{align}
\mathcal{A}^{(0)}_6|_{I_1I_25} &= \sum_{h=\sigma} \mathcal{A}^{(0)}_4\left({\bf I_1},{\bf I_2}, \hat 5^{\pm}, \hat P^{\sigma}\right) 
	{i\over s_{I_1I_25}} \mathcal{A}^{(0)}_4\left({\bf J_1},{\bf J_2}, \hat 6^{+}, -\hat P^{-\sigma}\right),
\end{align}
requires the scaling from the four-point single-flavor amplitude. 
From $\hbar$ counting we find
\begin{align}
	\mathcal{A}_4^{(0) }\left({\bf I}_1,{\bf I}_2, \hat { P}^{\pm},\hat 5^{+}\right) \stackrel{\hbar \to 0}{\sim} \hbar^{-1} \; .
\end{align}
And as the massless poles contributes $\hbar^{-2}$, the massless factorizations manifestly scale as
\begin{align}
\mathcal{A}^{(0) }_6|_{I_1I_25} \stackrel{\hbar \to 0}{\sim} \hbar^{-4} .
\end{align}
Thus we conclude that 
\begin{align}
	\mathcal A_6^{(0) }( {\bf 1}^A, {\bf 2}^B, {\bf 3}^A,{\bf 4}^B , 5^{\pm},6^+) \stackrel{\hbar \to 0}{\sim} \hbar^{-4}.
\end{align}

We expect similar arguments to hold at higher points, which would imply that the general scaling of the $(n+4)$-point amplitude is
\begin{align}
	\mathcal A_{4+n}^{(0) } \stackrel{\hbar \to 0}{\sim} \hbar^{-3-{n\over 2}} \; .
	\label{eq:conjecture_npt}
\end{align}

\subsection{Coherence of the final radiative state}

Using the classical scaling discussed in equations \eqref{eqn:momenta_resc1}-\eqref{eqn:coupling_resc}, we can rewrite the graviton emission probability in our problem as
\begin{align}
P_n^{\lambda} &=  \frac{1}{n!} \sum_{\sigma_1,...,\sigma_n = \pm} \Big\langle\hspace{-3pt}\Big\langle \hbar^{4 + 2 n} \int_{\bar{\lambda}} \prod_{i=1}^n d \Phi(\bar{k}_i) \int \frac{d^4 \bar{q}}{(2 \pi)^4}  \delta\left(2 p_{A} \cdot \bar{q}\right) \delta\left(2 p_{B} \cdot \bar{q}\right) \Theta\left(p_{A}^{0}+\hbar \frac{\bar{q}^{0}}{2}\right) \Theta\left(p_{B}^{0}-\hbar \frac{\bar{q}^{0}}{2}\right) \nonumber \\
   &\times \int d^4 \bar{w}_1 d^4 \bar{w}_2  e^{-i b \cdot \bar{q}}  \delta^{(4)}\left(\bar{w}_{1}+\bar{w}_{2}+\sum_{i=1}^n \bar{k}_i \right) \Theta\left(p_A^{0}+\hbar \bar{w}_1^{0}- \hbar \frac{\bar{q}^0}{2}\right) \Theta\left(p_B^{0}+\hbar \bar{w}_2^{0}+ \hbar \frac{\bar{q}^0}{2}\right) \nonumber \\
   &\times \delta\left(2 p_A \cdot \bar{w}_1+\hbar \bar{w}_1^{2} - \hbar \bar{q} \cdot \bar{w}_1\right)   \delta\left(2 p_B \cdot \bar{w}_2+\hbar \bar{w}_2^{2} + \hbar \bar{q} \cdot \bar{w}_2\right)  \nonumber \\
&\times \mathcal{A}_{n+4} \left(p_A - \hbar \frac{\bar{q}}{2}, p_B + \hbar \frac{\bar{q}}{2} \rightarrow p_A+\hbar \bar{w}_{1}- \hbar \frac{\bar{q}}{2}, p_B+\hbar \bar{w}_{2}+ \hbar \frac{\bar{q}}{2}, \hbar \bar{k}_1^{\sigma_1},...,\hbar \bar{k}_n^{\sigma_n}\right)  \nonumber \\
&\hspace{25pt} \times \mathcal{A}^{*}_{n+4}\left(p_A+\hbar \frac{\bar{q}}{2}, p_B-\hbar \frac{\bar{q}}{2} \rightarrow p_A+\hbar \bar{w}_{1}- \hbar \frac{\bar{q}}{2}, p_B+\hbar \bar{w}_{2}+ \hbar \frac{\bar{q}}{2}, \hbar \bar{k}_1^{\sigma_1},...,\hbar \bar{k}_n^{\sigma_n}\right) \Big\rangle\hspace{-3pt}\Big\rangle .
   \label{eqn:probabilityclass}
\end{align}
The leading order contribution in the classically relevant region is 
\begin{align}
P_n^{\lambda} &=  \frac{1}{n!} \sum_{\sigma_1,...,\sigma_n = \pm} \Big\langle\hspace{-3pt}\Big\langle \hbar^{4 + 2 n} \int_{\bar{\lambda}} \prod_{i=1}^n d \Phi(\bar{k}_i) \int \frac{d^4 \bar{q}}{(2 \pi)^4}  \delta\left(2 p_{A} \cdot \bar{q}\right) \delta\left(2 p_{B} \cdot \bar{q}\right) \Theta\left(p_{A}^{0}\right) \Theta\left(p_{B}^{0}\right) \nonumber \\
   &\times \int d^4 \bar{w}_1 d^4 \bar{w}_2  e^{-i b \cdot \bar{q}}  \delta^{(4)}\left(\bar{w}_{1}+\bar{w}_{2}+\sum_{i=1}^n \bar{k}_i \right) \delta\left(2 p_A \cdot \bar{w}_1\right) \delta\left(2 p_B \cdot \bar{w}_2\right) \nonumber \\
&\times \mathcal{A}_{n+4} \left(p_A - \hbar \frac{\bar{q}}{2}, p_B + \hbar \frac{\bar{q}}{2} \rightarrow p_A+\hbar \bar{w}_{1}- \hbar \frac{\bar{q}}{2}, p_B+\hbar \bar{w}_{2}+ \hbar \frac{\bar{q}}{2}, \hbar \bar{k}_1^{\sigma_1},...,\hbar \bar{k}_n^{\sigma_n}\right)  \nonumber \\
&\hspace{25pt} \times \mathcal{A}^{*}_{n+4}\left(p_A+\hbar \frac{\bar{q}}{2}, p_B-\hbar \frac{\bar{q}}{2} \rightarrow p_A+\hbar \bar{w}_{1}- \hbar \frac{\bar{q}}{2}, p_B+\hbar \bar{w}_{2}+ \hbar \frac{\bar{q}}{2}, \hbar \bar{k}_1^{\sigma_1},...,\hbar \bar{k}_n^{\sigma_n}\right) \Big\rangle\hspace{-3pt}\Big\rangle .
   \label{eqn:probabilityclass2}
\end{align}
It is the scaling of the \textit{energy} of the emitted radiation that determines if the amplitude contribution is classical or quantum, and in the following we take this as a guiding principle. The expectation value of the energy operator is given by the same unitarity cuts appearing in the mean of the graviton particle distribution, but weighted in the phase space integration by an energy factor $E_j := \hbar \omega_j$ for each of the emitted gravitons. The scaling in the classical limit has to be such that the total energy carried by the emitted gravitons, i.e.\ by the classical gravitational wave,
\begin{align}
   E^{\mathrm{cl}} &= \sum_{n=1}^{\infty} \frac{1}{n!} \sum_{\sigma_1,...,\sigma_n = \pm} \Big\langle\hspace{-3pt}\Big\langle \hbar^{5 + 2 n} \int \prod_{i=1}^n d \Phi(\bar{k}_i) \int \frac{d^4 \bar{q}}{(2 \pi)^4}  \delta\left(2 p_{A} \cdot \bar{q}\right) \delta\left(2 p_{B} \cdot \bar{q}\right) \nonumber \\
   &\times \int d^4 \bar{w}_1 d^4 \bar{w}_2  e^{-i b \cdot \bar{q}}  \delta^{(4)}\left(\bar{w}_{1}+\bar{w}_{2}+\sum_{i=1}^n \bar{k}_i \right) \delta\left(2 p_A \cdot \bar{w}_1\right) \delta\left(2 p_B \cdot \bar{w}_2\right) \left(\sum_{j=1}^n \omega_j\right)\nonumber \\
&\times \mathcal{A}_{n+4} \left(p_A - \hbar \frac{\bar{q}}{2}, p_B + \hbar \frac{\bar{q}}{2} \rightarrow p_A+\hbar \bar{w}_{1}- \hbar \frac{\bar{q}}{2}, p_B+\hbar \bar{w}_{2}+ \hbar \frac{\bar{q}}{2}, \hbar \bar{k}_1^{\sigma_1},...,\hbar \bar{k}_n^{\sigma_n}\right)  \nonumber \\
&\hspace{25pt} \times \mathcal{A}^{*}_{n+4}\left(p_A+\hbar \frac{\bar{q}}{2}, p_B-\hbar \frac{\bar{q}}{2} \rightarrow p_A+\hbar \bar{w}_{1}- \hbar \frac{\bar{q}}{2}, p_B+\hbar \bar{w}_{2}+ \hbar \frac{\bar{q}}{2}, \hbar \bar{k}_1^{\sigma_1},...,\hbar \bar{k}_n^{\sigma_n}\right) \Big\rangle\hspace{-3pt}\Big\rangle ,
\end{align}
is finite in the classical limit. While each separate probability of the emission of $n$ gravitons \eqref{eqn:probabilityclass2} is infrared divergent when $\lambda \to 0$, in this paper we are interested only in the deviation from a Poissonian distribution in the $\hbar \to 0$ limit,
\begin{align}
\lim_{\hbar \to 0} \hbar \,\Delta_{\textrm{out}} = \lim_{\hbar \to 0} \hbar \,(\Sigma_{\textrm{out}}^{\lambda} - \mu_{\textrm{out}}^{\lambda}) .
\end{align}
As we have shown in section \ref{sec:gravstats}, this is an infrared-safe quantity. A naive power counting in $\hbar$ from Feynman diagrams for the five-point and six-point tree gives a series expansion starting with the following types of terms,
\begin{align}
\mathcal{A}_{5}^{(0)} &\left(p_A - \hbar \frac{\bar{q}}{2}, p_B + \hbar \frac{\bar{q}}{2}, p_A+\hbar \bar{w}_{1}- \hbar \frac{\bar{q}}{2}, p_B+\hbar \bar{w}_{2}+ \hbar \frac{\bar{q}}{2}, \hbar \bar{k}_1^{\sigma_1}\right) \nonumber \\
&\hspace{180pt}= \frac{C^{\mathcal{A}_5^{(0)}}_1}{\hbar^{\frac{9}{2}}} +  \frac{C^{\mathcal{A}_5^{(0)}}_2}{\hbar^{\frac{7}{2}}} + \mathcal{O} \left(\frac{1}{\hbar^{\frac{5}{2}}}\right), \nonumber \\ 
\mathcal{A}_{6}^{(0)} &\left(p_A - \hbar \frac{\bar{q}}{2}, p_B + \hbar \frac{\bar{q}}{2}, p_A+\hbar \bar{w}_{1}- \hbar \frac{\bar{q}}{2}, p_B+\hbar \bar{w}_{2}+ \hbar \frac{\bar{q}}{2}, \hbar \bar{k}_1^{\sigma_1} , \hbar \bar{k}_2^{\sigma_2}\right) \nonumber \\
&\hspace{180pt}= \frac{C^{\mathcal{A}_6^{(0)}}_1}{\hbar^{6}} +  \frac{C^{\mathcal{A}_6^{(0)}}_2}{\hbar^{5}} +  \frac{C^{\mathcal{A}_6^{(0)}}_3}{\hbar^{4}} + \mathcal{O} \left(\frac{1}{\hbar^{3}}\right),
\end{align}
but as we have seen in the preceding subsections, it turns out that some of the lower-order terms are zero,
\begin{align}
C^{\mathcal{A}_5^{(0)}}_1 &= 0, \qquad C^{\mathcal{A}_5^{(0)}}_2 \neq 0, \nonumber \\ 
C^{\mathcal{A}_6^{(0)}}_1 &= C^{\mathcal{A}_6^{(0)}}_2 = 0, \qquad C^{\mathcal{A}_6^{(0)}}_3 \neq 0 .
\end{align}
The cancellation of the leading term in the $\hbar$ expansion was shown already in \cite{Luna2018} for $\mathcal{A}^{(0)}_5$, but the new result here is the double cancellation of two leading terms in the $\hbar$ expansion for $\mathcal{A}_6$.\footnote{A similar result has been obtained in scalar QED in \cite{Cristofoli:2021jas}.} This has physical consequences, as we have seen: we find that
\begin{align}
\lim_{\hbar \to 0} \hbar  P_1^{(0,0)} \sim \hbar^{0}, \qquad \lim_{\hbar \to 0} \hbar   \, P_2^{(0,0)}  = 0,
\end{align}
where for simplicity we have kept the powers of $\hbar$ coming from the coupling in \eqref{eqn:coupling_resc} implicit inside the probabilities.\footnote{Alternatively, we should have written 
\begin{align*}
\lim_{\hbar \to 0} \hbar \left(\frac{G^3}{\hbar^3}\right)  P_1^{(0,0)} \sim \hbar^{0}, \qquad \lim_{\hbar \to 0} \hbar  \left(\frac{G^4}{\hbar^4}\right) \, P_2^{(0,0)}  = 0.
\end{align*}
We have decided to avoid this cumbersome notation here.} This will be assumed for all the rest of our arguments in this section.

Therefore, while the 5-point tree-level amplitude gives a classical contribution to classical radiative observables, the 6-point tree-level amplitude gives a ``quantum'' contribution
\begin{align}
\lim_{\hbar \to 0} \hbar \,\Delta_{\textrm{out}} \Big|_{\mathcal{O}(G^4)} = 0 ,
\end{align}
which proves that we can describe the final semiclassical radiation state as a coherent state at least up to order $\mathcal{O}(G^4)$ for classical radiative observables.  

\subsection{Classical relations for unitarity cuts from all-order coherence}

Assuming coherence to all orders in perturbation theory implies a set of (integral) relations between loop and tree amplitudes with emission of gravitons. 

For example, we expect that unitarity cuts involving tree-level amplitudes with two or more gravitons emitted, and their conjugates, would give vanishing contributions in the classical limit. The reason is that having a coherent state as an exact final semiclassical state for the radiation would imply that all the gravitons emitted are uncorrelated. Indeed, our conjectural classical scaling for tree-level amplitudes in \eqref{eq:conjecture_npt}, 
\begin{align}
\mathcal{A}^{(0) }_{n+4}(\phi_A \phi_B \to \phi_A \phi_B h_1 h_2 .....h_n) \sim \hbar^{-3 -\frac{n}{2}}, 
\label{eq:scaling_npt}
\end{align}
would imply that
\begin{align}
\lim_{\hbar \to 0} \, \hbar  P_n^{(0,0)} = 0 \qquad \textrm{for~} n \geq 2.
\label{eq:P00_quantum}
\end{align}
This follows directly from our main result \eqref{eqn:main_result_Deltam}. While the expansion of $\Gamma_{\textrm{out}}^{(n),\lambda}$ starts at order $G^{2 + n}$, the lowest order contribution to $(\mu_{\textrm{out}}^{\lambda})^n$ is of order $G^{2 n + n}$: clearly then for $n \geq 2$ the equation \eqref{eq:P00_quantum} must  hold, as a simple consequence of coherence. 

\begin{figure}[h]
\centering
\includegraphics[scale=0.65]{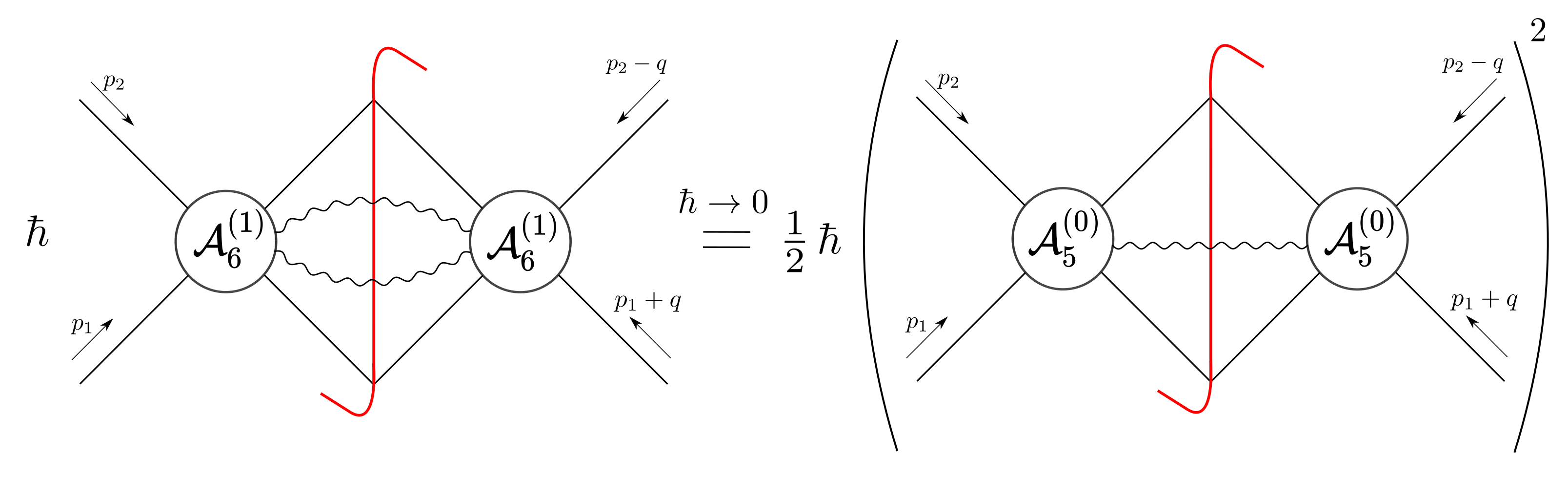}
\caption{If the final graviton particle distribution is Poissonian, this implies non-trivial classical relations between cut contributions of classical amplitudes with more than one graviton emitted to the cuts of the 5-point amplitude $\mathcal{A}_5({\bf 1}^A,{\bf 2}^B,{\bf 3}^A,{\bf 4}^B,5^{\sigma_1})$.}
\label{fig:classical_cut_relation}
\end{figure}

In order to make definite statements about the probabilities at higher orders, we need to combine them at a given loop order, so let us define
\begin{align}
P_{n}^{(L)} := \sum_{j=0}^{L} P_{n}^{(j, L-j)}.
\label{eq:homogeneous}
\end{align}
Once we feed \eqref{eq:homogeneous} and \eqref{eq:P00_quantum} into the constraints given by considering the factorial moments
\begin{align}\label{eq:coherence-Delta}
\lim_{\hbar \to 0} \hbar \, \Delta_{\textrm{out}}^{(j)} = 0 \qquad  \textrm{for~} j \geq 2,
\end{align}
we can conclude that
\begin{align}
\lim_{\hbar \to 0} \, \hbar P_n^{(L)} = \lim_{\hbar \to 0} \, \hbar \left(\sum_{j=0}^{L} P_{n}^{(j, L-j)}\right) = 0, \qquad \textrm{for~} L < 2(n-1) .
\label{eq:constraint_diag}
\end{align}
This is the loop-level generalization of \eqref{eq:P00_quantum}, which is essentially saying that coherence implies that classically we can only have, at a given loop order $L_1+L_2$, contributions from product of amplitudes with $n \leq (L_1+L_2)/2+1$ external gravitons.
Therefore if we consider the expansion of $\Delta_{\textrm{out}}^{(m)}$ with $m=2,3,4$ that we found in (\ref{eqn:Delta_expansion}), (\ref{eqn:Delta3_expansion}), and (\ref{eqn:Delta4_expansion}),
\allowdisplaybreaks
\begin{align}
\Delta^{(2)}_{\textrm{out}}  &= 2 G^4 P_2^{(0,0)} + 6 G^5 P_3^{(0,0)} + 12 G^6 P_4^{(0,0)} + 20 G^7 P_5^{(0,0)}  \nonumber \\
& + G^5 (2 P_2^{(1,0)} + 2 P_2^{(0,1)} ) + G^6 ( 2 P_2^{(2,0)} + 2 P_2^{(0,2)} + 6 P_3^{(1,0)} + 6 P_3^{(0,1)} ) \nonumber \\
& + G^7 (2 P_2^{(3,0)} + 2 P_2^{(0,3)} + 6 P_3^{(2,0)} + 6 P_3^{(0,2)}+ 6 P_3^{(1,1)} + 12 P_4^{(1,0)} + 12 P_4^{(0,1)} - 4 P_1^{(0,0)} P_2^{(0,0)}) \nonumber \\
& + \Big[G^6 (2 P_2^{(1,1)} - (P_1^{(0,0)})^2) + G^7 (2 P_2^{(1,2)} + 2 P_2^{(2,1)}- 2 P_1^{(0,1)} P_1^{(0,0)} - 2 P_1^{(1,0)} P_1^{(0,0)})\Big],\nonumber \\ \nonumber\\ \nonumber \\
\Delta^{(3)}_{\textrm{out}}  &= 6 G^5 P_3^{(0,0)} + 24 G^6 P_4^{(0,0)} + 60 G^7 P_5^{(0,0)} \nonumber \\
& + G^6 (6 P_3^{(1,0)} + 6 P_3^{(0,1)} ) + G^7 (6 P_3^{(0,2)} + 6 P_3^{(2,0)} + 6 P_3^{(1,1)} + 24 P_4^{(1,0)} + 24 P_4^{(0,1)} ), \nonumber \\
\Delta^{(4)}_{\textrm{out}}  &= 24 G^6 P_4^{(0,0)} + 120 G^7 P_5^{(0,0)} \nonumber \\
& + G^7 (24 P_4^{(1,0)} + 24 P_4^{(0,1)} )\, ,
\label{eqn:Deltaset_expansion}
\end{align}
we get, after imposing all the constraints \eqref{eq:P00_quantum} and \eqref{eq:constraint_diag} in the expansion in the coupling,
\begin{align}
\lim_{\hbar \to 0} \hbar \Delta^{(2)}_{\textrm{out}}  &= \lim_{\hbar \to 0} \hbar  \left(G^6 ( 2 P_2^{(2,0)} + 2 P_2^{(0,2)}) \right)\nonumber \\
&+ \lim_{\hbar \to 0} \hbar  \left(G^7 (2 P_2^{(3,0)} + 2 P_2^{(0,3)} + 6 P_3^{(2,0)} + 6 P_3^{(0,2)}+ 6 P_3^{(1,1)}\right)\nonumber \\
& + \lim_{\hbar \to 0} \hbar  \Big[G^6 (2 P_2^{(1,1)} - (P_1^{(0,0)})^2) + G^7 (2 P_2^{(1,2)} + 2 P_2^{(2,1)} - 2 P_1^{(0,1)} P_1^{(0,0)} - 2 P_1^{(1,0)} P_1^{(0,0)})\Big],\nonumber \\
\lim_{\hbar \to 0} \hbar \Delta^{(3)}_{\textrm{out}}  &= \lim_{\hbar \to 0} \hbar  \left(G^7 (6 P_3^{(0,2)} + 6 P_3^{(2,0)} + 6 P_3^{(1,1)}) \right), \nonumber \\
\lim_{\hbar \to 0} \hbar \Delta^{(4)}_{\textrm{out}}  & = 0 .
\label{eqn:Deltaset_expansion2}
\end{align}
Assuming (\ref{eq:coherence-Delta}) as a consequence of coherence, we can now impose
\begin{align}
\lim_{\hbar \to 0} \hbar \Delta^{(3)}_{\textrm{out}} = \lim_{\hbar \to 0} \hbar  \left(G^7 (6 P_3^{(0,2)} + 6 P_3^{(2,0)} + 6 P_3^{(1,1)}) \right) = 0,
\label{eqn:P311eq}
\end{align}
which implies for $\Delta^{(2)}_{\textrm{out}}$
\begin{align}
\lim_{\hbar \to 0} \hbar \Delta^{(2)}_{\textrm{out}}  &= \lim_{\hbar \to 0} \hbar  \left(G^6 ( 2 P_2^{(2,0)} + 2 P_2^{(0,2)}) +G^7 (2 P_2^{(3,0)} + 2 P_2^{(0,3)})\right)\nonumber \\
& + \lim_{\hbar \to 0} \hbar  \Big[G^6 (2 P_2^{(1,1)} - (P_1^{(0,0)})^2) + G^7 (2 P_2^{(1,2)} + 2 P_2^{(2,1)} - 2 P_1^{(0,1)} P_1^{(0,0)} - 2 P_1^{(1,0)} P_1^{(0,0)})\Big].
\label{eqn:finalDelta2exp}
\end{align}
We see now that the contributions in the first line manifestly involve the six-point tree amplitude and six-point loop amplitudes. We expect from \eqref{eq:constraint_diag}, and also from the uncertainty principle \cite{Cristofoli:2021jas}, that the amplitudes ${\mathcal A}_n^{(L)}$ with $n \geq L+2$ always give a vanishing contribution to observables in the classical limit because they do not contribute to the classical field. In particular, this applies to the six-point tree-level amplitude ${\mathcal A}_2^{(0)}$ and higher-point tree-level amplitudes. But we cannot prove this directly from the coherence property, so we therefore assume that this is the case. We can then conjecture that
\begin{align}
\lim_{\hbar \to 0} \, \hbar P_{n}^{(L,0)}= \lim_{\hbar \to 0} \, \hbar P_{n}^{(0,L)} &= 0 \qquad \textrm{for~} n \geq 2 \,,
\label{eq:superclassicalzero}
\end{align}
which is equivalent to saying that the leading classical term in the expansion of the $L$-loop $(4+n)$-point amplitudes will not conspire with the quantum $\hbar$ scaling of the $(4+n)$-point tree amplitude to give a classical contribution. It would be nice to have a direct check of \eqref{eq:superclassicalzero} and its higher order generalizations. A first consequence of \eqref{eqn:P311eq} and \eqref{eq:superclassicalzero} is 
\begin{align}
\lim_{\hbar \to 0} \, \hbar P_{3}^{(1,1)} = 0 ,
\label{eq:sevenpoint-tree}
\end{align}
which is equivalent to the statement that the seven-point one-loop amplitude is classically suppressed.  More generally, from the equations \eqref{eq:coherence-Delta} and \eqref{eq:superclassicalzero} a very interesting set of relations follow directly for successive values of the loop order and the order in the coupling $G$, beginning with the following:
\begin{align}
\hbar P_2^{(1,1)} &\stackrel{\hbar \to 0}{=} \hbar \frac{1}{2} (P_1^{(0,0)})^2 \nonumber \\
\hbar (P_2^{(1,2)} + P_2^{(2,1)}) &\stackrel{\hbar \to 0}{=} \hbar (P_1^{(0,1)} P_1^{(0,0)} + P_1^{(1,0)} P_1^{(0,0)})\nonumber \\
	\hbar (P_2^{(1,3)}+P_2^{(2,2)} + P_2^{(3,1)}) &\stackrel{\hbar \to 0}{=} \hbar ({1\over 2}
	(P_1^{(1,1)})^2  + P_1^{(1,0)} P_1^{(1,1)})\nonumber \\
	\hbar (P_3^{(1,3)}+P_3^{(2,2)} + P_3^{(3,1)}) &\stackrel{\hbar \to 0}{=} 
	 \hbar \frac{1}{6}(P_1^{(0,0)})^3\;.
\label{eq:classical_relations}
\end{align}
The similarity between the top and bottom line emerges from the combinatorics of the power matching. This type of relation has a straightforward generalisation to families relating $n$-graviton amplitudes at $L_1+L_2=2n-2+k$ for $k=0,1,2,\dots$, which 
can be generated from the matching of orders of $G$ in the factorial moments. Those relations have the common feature that they relate particular combinations of unitarity cuts involving more than one graviton emitted at higher loops to other unitarity cuts involving the 5-point amplitude at a lower loop level. We have represented the simplest of these relations, involving the one-loop amplitude with two gravitons emitted and the tree-level amplitude $\mathcal{A}^{(0)}_5({\bf 1}^A,{\bf 2}^B,{\bf 3}^A,{\bf 4}^B,5^{\sigma_1})$ in Fig.~\ref{fig:classical_cut_relation}. 

The outcome of this section is that we have strong evidence that the fundamental data to describe the final semiclassical state are encoded in the 5-point amplitude at all orders in the coupling constant, providing that  \eqref{eq:superclassicalzero} and its higher order generalizations hold. All the higher-multiplicity amplitudes are either suppressed in the classical regime, or related to the 5-point amplitude by a classical relation. This suggests, purely from the S-matrix perspective, that we can describe the radiation in the two-body problem entirely with a coherent state where the 5-point amplitude $\mathcal{A}_5(\phi_A \phi_B \to \phi_A \phi_B h_1)$ plays an essential role, as suggested in \cite{Cristofoli:2021jas}. 

\section{Conclusion}
\label{sec:conclusion}

In an effective field theory approach to the scattering of compact bodies in GR, we can reduce the problem to considering a pair of minimally coupled massive scalar particles interacting with gravitons in perturbation theory. The KMOC formalism provides a rigorous framework to take the classical limit of quantum scattering amplitudes for massive particles \cite{Kosower:2018adc}, and this was recently extended to the scattering of waves by using coherent states \cite{Cristofoli:2021vyo}. Essentially, this is an on-shell classical limit of the in-in formalism at zero temperature which is built into the standard framework of quantum field theory. For the two-body problem in general relativity, the incoming KMOC state for two massive particles is a pure state. The unitarity of the S-matrix then dictates that ingoing pure states are mapped to outgoing pure states, and classical pure states for the radiation field are known to be described exactly by one coherent state \cite{HILLERY1985409}. 

In this paper, we have found evidence of this fact by studying scattering amplitudes with external gravitons. In particular, we have studied the properties of the final graviton particle distribution using the Glauber-Sudarshan representation \cite{Glauber:1963fi,Glauber:1963tx,Sudarshan:1963ts}. We have considered the mean, the variance and higher-order factorial moments of the distribution by taking the appropriate expectation values of the graviton number operator in the KMOC formalism.  Since coherent states are characterized by exact Poissonian statistics, the deviation from a coherent state structure is conveniently parametrized by the difference $\Delta^{(m)}$ between the factorial moments and the expected value for a Poisson distribution. Given that zero-energy gravitons in our problem obeys exactly a Poissonian distribution, $\Delta$ is infrared finite.  In the perturbative expansion, we proved that the leading contribution is related to the unitarity cut involving the six-point tree amplitude $\mathcal{A}^{(0)}_6(\phi_A \phi_B \to \phi_A \phi_B h_1 h_2)$ and its conjugate. This is expected, since the deviation from coherence has to come from the correlation between graviton emissions.\footnote{Or from a non-zero entanglement, along the lines of \cite{Aoude:2020mlg}.}

The crucial problem is therefore to compute this tree-level amplitude and and its classical scaling. To do that, we developed two new approaches. First, we extend the Cheung-Remmen parametrization of the pure Einstein-Hilbert action to include minimally coupled scalars. The obtained Feynman rules are very compact, and we were able to compute analytically the full amplitude with 68 diagrams. Second, we constructed on-shell recursion relations for the case of tree-level amplitudes with two different massive particles flavors coupled to gravity: a new ``equal-mass shift'' is used to construct the 5-point amplitude $\mathcal{A}^{(0)}_5(\phi_A \phi_B \to \phi_A \phi_B h_1)$ and the standard BCFW shift was then used to compute the 6-point amplitude $\mathcal{A}^{(0)}_6(\phi_A \phi_B \to \phi_A \phi_B h_1 h_2)$. While the large $z-$scaling behavior is non-trivial, a direct calculation shows that the boundary terms vanish, justifying  our approach. We found perfect agreement between the two approaches, and we also agree with known results in the literature for the 5-point amplitude \cite{Luna2018}.

Regarding the classical limit, we label as ``classical'' the amplitudes which give a contribution to the total energy emitted in the classical limit in the KMOC formalism. The unitarity cuts appearing in such an expectation value are the same as for the probability of graviton emission, and therefore the $\hbar$ scaling of the amplitudes appearing in those cuts determines whether we get a classical or a quantum contribution. It is known that a naive $\hbar$ power counting does not give the correct answer, as this was already pointed out in \cite{Luna2018} for the 5-point tree by doing an equivalent large mass expansion. Here we showed this in a manifestly gauge-invariant way in the spinorial formalism, by defining suitable kinematic variables which have a well-defined $\hbar$ expansion. We confirmed the classical result for $\mathcal{A}^{(0)}_5$ obtained in \cite{Luna2018}, and we found that the six-point amplitude $\mathcal{A}^{(0)}_6$ gives a purely quantum contribution. A BCFW-like argument suggests that $\mathcal{A}^{(0)}_n \stackrel{\hbar \to 0}{\sim} \hbar^{-3 -\frac{n}{2}}$, which would mean that tree-level amplitudes with an higher number of emitted gravitons should also give a quantum contribution. This result also resonates with some conjectural classical relations that we found between unitarity cuts of scalar graviton amplitudes, which point towards a characterization of the coherent state only in terms of the (all order) 5-point amplitude data. While this is often implicitly assumed in some wordline descriptions \cite{Mogull:2020sak,Jakobsen:2021lvp,Jakobsen:2021smu}, our result provides a direct justification from the S-matrix perspective. Further developments along these lines have been pursued in \cite{Cristofoli:2021jas}.\footnote{Some of the ideas in this direction have been presented at the Paris-Saclay AstroParticle Symposium 2021 by P.~Di Vecchia, C.~Heissenberg, R.~ Russo and G.~Veneziano in a series of seminars \cite{carlosTalk,rodolfosTalk,gabrielesTalk}.}

Our work has further connections in several other directions. For example, in the case of quantum field theory with external sources, unitarity cuts involving vacuum diagrams have been related to the Abramovsky-Gribov-Kancheli (AGK) cancellation in the context of context of reggeon field theory models in \cite{Abramovsky:1973fm,Gelis:2006cr,Gelis:2006yv}. There it was shown that a Poissonian distribution of the cut reggeons naturally explain the AGK cancellation, and this actually inspired part of the ideas developed in this work. Furthermore, the set of infinite amplitude relations we found must have some overlap with the ones related to soft theorems \cite{Bautista:2021llr,Gonzo:2020xza}, which in general are also valid beyond the classical regime. It would be nice to make this connection more precise. Finally, we have only discussed the classical perturbative long-distance regime of the scattering, but quantum and classical non-perturbative effects can make radiation incoherent and will introduce correlations between the waveform detected at different locations. It would be interesting to explore this further.

We conclude with some open questions. First, it is known that the classical description breaks down at sufficiently high energies because of quantum radiation reaction effects, which ultimately make the emitted gravitons interfere with each other \cite{Herrmann:2021tct,Herrmann:2021lqe,
Ciafaloni:2015xsr,Ciafaloni:2018uwe}. This is actually important to have a consistent resummation of radiation reaction effects, and perhaps a simpler setup where analytic calculations are possible at very high orders -- like working in a fixed background -- can give us some useful lessons in this direction \cite{DiPiazza:2010mv,Dinu:2012tj,Seipt:2012tn,Dinu:2015aci,Ilderton:2017xbj,Blackburn:2018sfn,Torgrimsson:2021wcj,Adamo:2021jxz}. Second, we are still lacking a rigorous proof of the general validity of on-shell recursion techniques in the case of a pair of massive particles minimally coupled with gravity, which would be helpful in establishing rigorously the all multiplicity tree-level classical scaling discussed in this work. Finally, we have restricted our attention to point particles. But spin and tidal effects (and possibly other higher-dimensional operators) can also be relevant, and it is not clear if coherence will persist once those operators are added to the lagrangian.

\acknowledgments
We are grateful to Andrea Cristofoli, Nathan Moynihan, Donal O'Connell, Matteo Sergola, Alasdair Ross and Chris White for discussions and collaboration on related topics. We thank the Galileo Galilei Institute for Theoretical Physics (GGI) and the program ``Gravitational scattering, inspiral, and radiation'' for providing a stimulating environment where some of these ideas have been developed. R.G. would like to thank the Institut de Physique Théorique (IPhT) for the hospitality during the latest stages of the project and Fran\c{c}ois Gelis for interesting discussions on the topic. This project has received funding from the European Union's Horizon 2020 research and innovation program under the ERC Consolidator Grant No. 647356 ``CutLoops'' and the  Marie Sklodowska-Curie ITN No.764850 ``SAGEX'', and from Science Foundation Ireland through grant 15/CDA/3472.

\appendix

\section{KMOC formalism from the classical on-shell reduction of the in-in formalism}
\label{sec:appendix_SK}

\begin{figure}[h]
\centering
\includegraphics[scale=0.3]{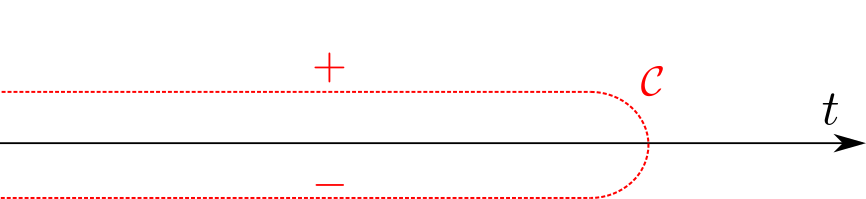}
\caption{The two branches of the Schwinger-Keldysh contour $\mathcal{C}$ run from above (+) to below (-) the real time axis.}
\label{fig:contour}
\end{figure}

This section is inspired by \cite{Gelis:2019yfm}. Without loss of generality, we will compute here perturbatively the in-in expectation value of the graviton number operator in pure Einstein gravity.\footnote{Later we will include matter coupled with gravity, in order to take the appropriate classical limit using the KMOC formalism.} Consider the expression
\begin{align}
\bra{0_{\text{in}}} \, a^{\dagger}_{\sigma}(k) a_{\sigma}(k) \ket{0_{\text{in}}}
\label{eqn:numb_oper_integrand}
\end{align}
purely from the Schwinger-Keldysh (SK) perspective,\footnote{This is also called the in-in formalism at zero temperature.} where $\ket{0_{\text{in}}}$ is the initial graviton state at $t=-\infty$, and $a^{\dagger}_{\sigma}(k)/a_{\sigma}(k)$ are the graviton creation/annihilation operators of a definite helicity $\sigma$. One can express \eqref{eqn:numb_oper_integrand} with the LSZ reduction as
\begin{align}
\bra{0_{\text{in}}} \, a^{\dagger}_{\sigma}(k) a_{\sigma}(k) \ket{0_{\text{in}}} =  \varepsilon_{\sigma}^{\mu \nu}(k) \varepsilon_{\sigma}^{\alpha \beta}(k) \int d^4 x \int d^4 y \, e^{i k \cdot (x - y) /\hbar} \DAlambert_x \DAlambert_y \bra{0_{\text{in}}} \, h_{\mu \nu}(x)  h_{\alpha \beta}(y) \ket{0_{\text{in}}}.
\end{align}
Notice that there is no (time) ordering in the correlator function. We now need to make contact with a generating functional to be able to compute this expression in perturbation theory. The idea is to introduce a new complex contour, called the Keldysh contour, which is made of two branches called $+$ and $-$ running parallel to the usual time axis (see Fig.~\ref{fig:contour}) and to formally double the set of fields $h^{(\pm)}$ involved in the path integral. Each copy of the fields will be labelled by the index $+$ or $-$ according to the branch of the contour $\mathcal{C}$ they belong to.

Using the interaction representation for the quantum fields, we can write\footnote{For simplicity we have suppressed the spacetime indices in the path integral variables and the boundary conditions of the path integral, which should force the state to be $\ket{0_{\text{in}}}$ at $t=-\infty$.}
\begin{align}
\bra{0_{\text{in}}} \, h_{\mu \nu}(x)  h_{\alpha \beta}(y) \ket{0_{\text{in}}} = \int \mathcal{D} h^{(+)} \mathcal{D} h^{(-)} \,  h^{(-)}_{\mu \nu}(x)  h^{(+)}_{\alpha \beta}(y) e^{i \int_{\mathbb{R} \times \mathbb{R}^3} d^4 x \left(\mathcal{L}^{(+)}_{\text{GR},int}[h^{(+)}] - \mathcal{L}^{(-)}_{\text{GR},int}[h^{(-)}] \right)} ,
\end{align}
where $\{\mathcal{L}^{(+)}_{\text{GR},int}[h^{(+)}],\mathcal{L}^{(-)}_{\text{GR},int}[h^{(-)}]\}$ is a set of two copies of the interaction lagrangian in the pure gravity theory where all the fields belong the same branch of the contour $\mathcal{C}$. At this point we can rewrite the initial expression as
\begin{align}
\bra{0_{\text{in}}} & \, a^{\dagger}_{\sigma}(k) a_{\sigma}(k) \ket{0_{\text{in}}} \nonumber \\
&=  \varepsilon_{\sigma}^{\mu \nu}(k) \varepsilon_{\sigma}^{\alpha \beta}(k) \int d^4 x \int d^4 y \, e^{i k \cdot (x - y) /\hbar} \DAlambert_x \DAlambert_y \bra{0_{\text{in}}} \, P  h^{(-)}_{\mu \nu}(x)  h^{(+)}_{\alpha \beta}(y) e^{i \int_{\mathbb{\mathcal{C}} \times \mathbb{R}^3} d^4 x \mathcal{L}_{\text{GR},int}[h])} \ket{0_{\text{in}}} ,
\label{eqn:num_operator_interaction}
\end{align}
where the ordering $P$ corresponds to 
\begin{align}
P h_{\mu \nu}(x)  h_{\alpha \beta}(y) = 
\begin{cases}
T h_{\mu \nu}(x)  h_{\alpha \beta}(y) & \quad \text{if} \qquad x_0,y_0 \in \mathcal{C}_+ \\
\bar{T} h_{\mu \nu}(x)  h_{\alpha \beta}(y) & \quad \text{if} \qquad x_0,y_0 \in \mathcal{C}_- \\
h_{\mu \nu}(x)  h_{\alpha \beta}(y) & \quad \text{if} \qquad x_0 \in \mathcal{C}_- , y_0 \in \mathcal{C}_+ \\
h_{\alpha \beta}(y) h_{\mu \nu}(x) & \quad \text{if} \qquad x_0 \in \mathcal{C}_+ , y_0 \in \mathcal{C}_- .
\end{cases}
\end{align}
We have therefore unified the treatment of the two time orderings in a compact way and have arrived at a simple path integral representation for the general in-in expectation value. Indeed, one can write a generating functional
\begin{align}
\mathcal{Z}^{\text{SK}}[j^{(+)},j^{(-)}] := \bra{0_{\text{in}}} e^{i \int_{\mathbb{\mathcal{C}} \times \mathbb{R}^3} d^4 x (\mathcal{L}_{\text{GR},int}[h] + j^{\mu \nu} h_{\mu \nu})} \ket{0_{\text{in}}}
\end{align}
in terms of which \eqref{eqn:num_operator_interaction} can be written as
\begin{align}
\bra{0_{\text{in}}} & \, a^{\dagger}_{\sigma}(k) a_{\sigma}(k) \ket{0_{\text{in}}} \nonumber \\
&=  \varepsilon_{\sigma}^{\mu \nu}(k) \varepsilon_{\sigma}^{\alpha \beta}(k) \int d^4 x \int d^4 y \, e^{i k \cdot (x - y) /\hbar} \DAlambert_x \DAlambert_y  \frac{\delta \mathcal{Z}^{\text{SK}}[j^{(+)},j^{(-)}] }{i \delta j^{\mu \nu, (+)}(x) i \delta j^{\alpha \beta, (-)}(y)} \Bigg|_{j^{(\pm)} = 0} .
\label{eqn:num_operator_interaction2}
\end{align}
The generic SK graviton propagator in the $(+)/(-)$ basis can then be written as
\begin{align}
\int d^4 x e^{i k \cdot x /\hbar} \bra{0_{\text{in}}} \, P  h^{(\eta)}_{\mu \nu}(x)  h^{(\eta')}_{\alpha \beta}(0) \ket{0_{\text{in}}} 
&=\begin{pmatrix}
      G_{\mu \nu \alpha \beta}^{++}(k) & G_{\mu \nu \alpha \beta}^{-+}(k) \\
      G_{\mu \nu \alpha \beta}^{+-}(k) & G_{\mu \nu \alpha \beta}^{--}(k) \\
\end{pmatrix} 
\nonumber \\
&=\begin{pmatrix}
      \frac{i }{\hbar^2 \bar{k}^2 + i \epsilon} & 2 \pi \theta(-\bar{k}_0) \delta(\hbar^2 \bar{k}^2)  \\
      2 \pi \theta(\bar{k}_0) \delta(\hbar^2 \bar{k}^2)  & -\frac{i}{\hbar^2 \bar{k}^2 - i \epsilon} \\
\end{pmatrix} \mathcal{P}_{\mu \nu \alpha \beta} ,
\end{align}
where $\eta,\eta'$ can take values $\pm 1$, and $\mathcal{P}_{\mu \nu \alpha \beta} := -\frac{1}{2}\left(\eta_{\mu \alpha} \eta_{\nu \beta}+\eta_{\mu \beta} \eta_{\nu \alpha}-\eta_{\mu \nu} \eta_{\alpha \beta}\right)$ in de Donder gauge. It is manifest that we can choose any basis for the SK formulation, for example the time-ordered/anti time-ordered (also called $(+)/(-)$) basis as in the previous calculations or the retarded/advanced basis, and the result will be independent of that choice. 

The direct connection with the standard Feynman integral perturbative expansion can be seen directly at the level of the generating functional. We can express the SK generating functional in terms of the Feynman generating functional and its conjugate 
\begin{align}
\mathcal{Z}^{\text{SK}}[j^{(+)},j^{(-)}] = e^{\int d^4 x d^4 y G^{\mu \nu \alpha \beta, + -}(x,y) \DAlambert_x \DAlambert_y \frac{\delta^2}{i \delta j^{\mu \nu,(+)}(x) i \delta j^{\alpha \beta,(-)}(y)} } \mathcal{Z}[j^{(+)}] \mathcal{Z}^*[j^{(-)}] .
\end{align}
Thanks to this equation, one can compute diagrams in the SK formalism by stitching together ordinary Feynman diagrams and their complex conjugates.

To make the connection with the KMOC formalism, we need to add matter coupled with gravity and to consider as our initial state $\ket{\psi_{\text{in}}}$. Essentially all the previous arguments can be generalized to extend the discussion for a correlator of a set of massive scalar and graviton fields. Then we have
\begin{align}
\bra{\psi_{\text{in}}} \, a^{\dagger}_{\sigma}(k) a_{\sigma}(k) \ket{\psi_{\text{in}}} =  \varepsilon_{\sigma}^{\mu \nu}(k) \varepsilon_{\sigma}^{\alpha \beta}(k) \int d^4 x \int d^4 y \, e^{i k \cdot (x - y) /\hbar} \DAlambert_x \DAlambert_y \bra{\psi_{\text{in}}} \, h_{\mu \nu}(x)  h_{\alpha \beta}(y) \ket{\psi_{\text{in}}},
\end{align}
and when we connect this with the interaction representation,
\begin{align}
\bra{\psi_{\text{in}}} \, P  h^{(-)}_{\mu \nu}(x)  h^{(+)}_{\alpha \beta}(y) e^{i \int_{\mathbb{\mathcal{C}} \times \mathbb{R}^3} d^4 x \mathcal{L}_{\text{GR+matter},\text{int}}[h])} \ket{\psi_{\text{in}}},
\end{align}
we must take the LSZ reduction for the massive external states with the appropriate KMOC wavefunction $\psi_A(p_1)$ and $\psi_B(p_2)$ as defined in \eqref{eqn:wavef_def},
\begin{align}
\int d \Phi(p_1) d \Phi(p_2) \, \psi_A(p_1) \psi_B(p_2) e^{i \frac{p_1 \cdot b}{\hbar}} \prod_{i=1}^2 \left[ \int d^4 x_i \, e^{i \frac{p_i \cdot x_i}{\hbar}} \left(\DAlambert_{x_i} + \frac{m_i^2}{\hbar^2}\right) \right] \langle ... \phi(x_1) \phi(x_2) ...\rangle ,
\end{align}
which in the limit $\hbar \to 0$ will effectively localize the massive particles on their classical trajectories characterized by a 4-velocity $v_A$ and $v_B$ and by an impact parameter $b^{\mu}$. 

\begin{figure}[h]
\centering
\includegraphics[scale=0.75]{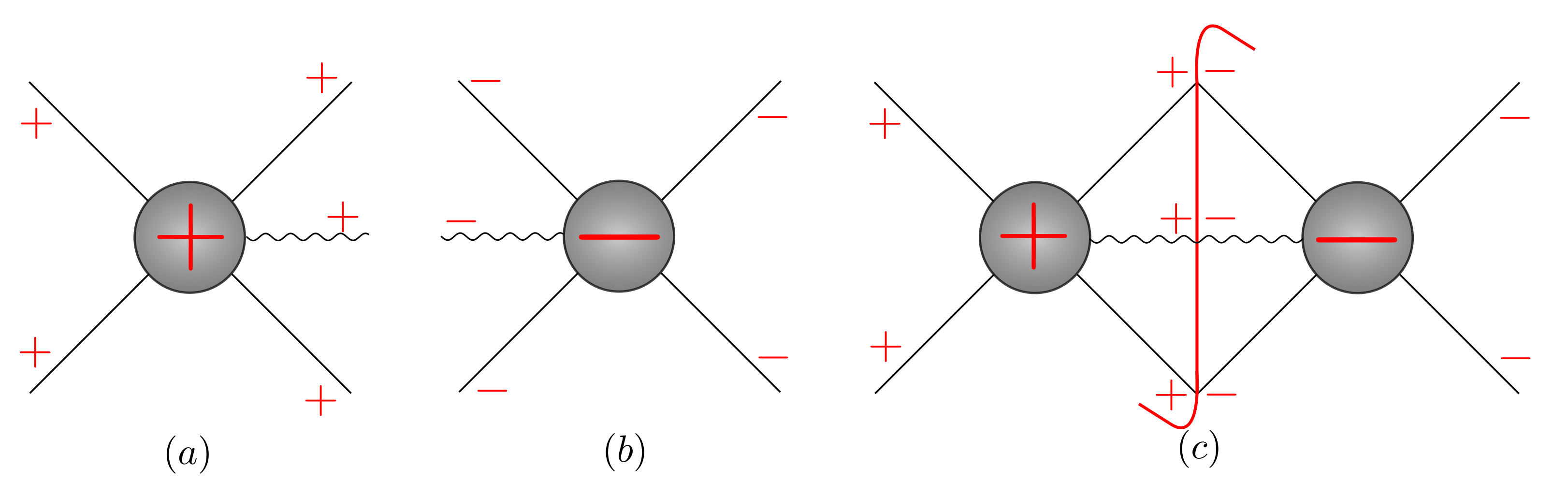}
\caption{The contributions of the type $(a)$ (resp.\ $(b)$) arise from purely time-ordered fields (resp.\ anti-time ordered) and correspond, under LSZ reduction for the external legs, to on-shell contributions which are linear in the amplitudes. On the other hand, terms of the type $(c)$ mix fields on different branches of the Schwinger-Keldysh contour, which corresponds in unitarity cut contributions between one amplitude and its conjugate in the on-shell formalism.}
\label{fig:Schwinger-Keldysh}
\end{figure}

The in-in formalism is a set of off-shell techniques in QFT which in principle can be used to compute the expectation value of any quantum field or polynomial thereof, including for example the stress tensor and its conserved charges. Here we have shown that taking an appropriate LSZ reduction on the external legs and using appropriate wavefunctions for the massive particles, we naturally obtain the KMOC formalism. Under LSZ reduction, the contraction arising from time-ordered $(+)$ or anti-time ordered $(-)$ correlators of fields $\{h_{\mu \nu},\phi\}$ in the Schwinger-Keldysh formalism maps to S-matrix elements (with the $+i \epsilon$ prescription) and their conjugates (with the $-i \epsilon$ prescription). Moreover, the contraction of fields belonging to different branches of the contour  ($(+)$ and $(-)$ or vice versa) gives the unitarity cut contributions. See Fig.~\ref{fig:Schwinger-Keldysh} for a pictorial representation of these different contributions. We hope that this will help to address some concerns raised by Damour in \cite{Damour:2019lcq,Damour:2020tta} on getting classical observables from scattering amplitudes with a definite $i \epsilon$ prescription.

This derivation gives some insight into the relation between the SK formalism and the KMOC formalism relevant to fully on-shell calculations, like the radiated energy, angular momentum, or more localized observables like the waveform and gravitational event shapes (essentially by considering only the on-shell radiative contribution of the fields arising in the large $r$ limit). But it also extends beyond this. In particular, it explains some recent derivations of off-shell metric configurations from ``amplitudes'' with one off-shell graviton leg \cite{Cristofoli:2020hnk}.
In that case one avoids taking the LSZ reduction of the graviton field whose expectation value is taken.
 A simple example is given by the (linearized) metric generated by on-shell matter particles coupled to gravity. For example, this justifies the results obtained in \cite{Cristofoli:2020hnk} for the derivation of gravitational shock wave configurations from the 3-point function with two massless on-shell scalars and one off-shell graviton. The same argument can be repeated for any other on-shell matter configuration coupled to one off-shell graviton, essentially making use of the (linearized) stress tensor \cite{Bjerrum-Bohr:2018xdl,Bautista:2021wfy}. Alternatively, one can work fully on-shell but in $(2,2)$ signature, as shown in \cite{Monteiro:2020plf,Crawley:2021auj,Guevara:2021yud}.

\section{Poissonian distributions and coherent states}
\label{sec:Poissonian}

The graviton coherent states introduced in the main text can be expanded in the Fock space basis of a definite number of gravitons,
\begin{align}
\ket{\alpha^{\sigma}} = \exp\left(-\frac{1}{2} \int d \Phi(k)  |\alpha^{\sigma}(k)|^2 \right) \sum_{n=0}^{\infty} \frac{1}{n!} \int \prod_{i=1}^n \left[ d \Phi(k_i) \alpha^{\sigma}(k_i) \right] \ket{k_1^{\sigma} ... k_n^{\sigma}},
\end{align}
and a direct calculation of the probability of detecting $n$ gravitons with helicity $\sigma'$ gives
\begin{align}
P_n^{\sigma'} :=& \delta_{\sigma \sigma'} \exp\left(-\int d \Phi(k)  |\alpha^{\sigma}(k)|^2 \right) \frac{1}{n!}  \left(\int d \Phi(k) |\alpha^{\sigma}(k)|^2 \right)^n,
\end{align}
which corresponds exactly to Poissonian statistics. A straightforward calculation of the mean and the variance in a coherent state gives
\begin{align}
\mu_{\alpha^{\sigma}} = \Sigma_{\alpha^{\sigma}} = \int d \Phi(k)  |\alpha^{\sigma}(k)|^2.
\end{align}
Poissonian statistics are equivalent to having a coherent state, as can be seen by computing $P_n^{\sigma'}$ for a generic probability distribution in the Glauber-Sudarshan representation,
\begin{align}
\Tr\left(P_n^{\sigma'}\right)_{\rho_{\text{radiation,GS}}} =& \sum_{\sigma=\pm} \int \mathcal{D}^2 \alpha^{\sigma} \, \mathcal{P}^{\sigma}(\alpha) \, P_n^{\sigma},
\end{align}
which requires $\mathcal{P}^{\sigma}(\alpha) = \delta^2(\alpha^{\sigma} - \alpha_{\star}^{\sigma})$ to match the Poissonian distribution.

In classical physics, however, we can have more general statistics for the classical radiation field. In particular, the variance of the distribution can be greater than the mean,\footnote{We expect the opposite inequality, i.e.\ $\mu_{\rho} > \Sigma_{\rho}$, to be relevant for purely quantum particle statistics. Indeed by using the Jensen inequality it is possible to prove that any mixture of coherent states will produce only super-Poissonian distributions.}
\begin{align}
\mu_{\rho} < \Sigma_{\rho},
\end{align}
which defines the so-called super-Poissonian statistics. This applies, for example, to thermal classical distributions. In our case, as discussed in the main text, the fact that we are working with pure states that are evolved with a unitary map suggests that all the classical states will have to obey the minimum uncertainty principle \cite{Kosower:2018adc}.





\bibliographystyle{JHEP-2}
\bibliography{references}





\end{document}